%% file: helga4.tex
\documentclass[preprint]{aastex}

\RequirePackage{graphicx,twoopt}

\defcitealias{2011arXiv1112.3348F}{Paper I} 	
\defcitealias{2012arXiv1204.0785S}{Paper II}
\defcitealias{2012ford}{Paper III}
\defcitealias{2008ApJ...675..330S}{S08}
\defcitealias{2007ApJ...654..240R}{R07}

\shorttitle{The distribution and properties of molecular clouds in M31}
\shortauthors{Kirk et al}

\begin{document}

\title{The Herschel\footnote{Herschel is an ESA space observatory with science instruments provided by European-led Principal Investigator consortia and with important participation from NASA.} Exploitation of Local Galaxy Andromeda (HELGA). VI. The distribution and properties of molecular cloud associations in M31} 

\author{
J.~M.~Kirk\altaffilmark{1,2}, 
W.~K.~Gear\altaffilmark{1},
J~ Fritz\altaffilmark{3},
M.~W.~L.~Smith\altaffilmark{1},
G.~Ford\altaffilmark{1},
M.~Baes\altaffilmark{3}, 
G.~J.~Bendo\altaffilmark{4},
I.~De~Looze\altaffilmark{3},
S.~A.~Eales\altaffilmark{1},
G.~Gentile\altaffilmark{3,5}, 
H.L.~Gomez\altaffilmark{1},
K.~Gordon\altaffilmark{6,3},	
B.~O'Halloran\altaffilmark{7},
S.~C.~Madden\altaffilmark{8},
J.~Roman~Duval\altaffilmark{6}, 
J.~Verstappen\altaffilmark{3},
S.~Viaene\altaffilmark{3}
A.~Boselli\altaffilmark{9},
A.~Cooray\altaffilmark{10},
V.~Lebouteiller\altaffilmark{8},
L.~Spinoglio\altaffilmark{11},
}
\altaffiltext{1}{Cardiff School of Physics and Astronomy, Cardiff University, Queens Buildings, The Parade, Cardiff, Wales, CF24 3AA, UK}
\altaffiltext{2}{Jeremiah Horrocks Institute, University of Central Lancashire, Preston PR1 2HE, UK}
\altaffiltext{3}{Sterrenkundig Observatorium, Universiteit Gent, Krijgslaan 281 S9, B-9000 Gent, Belgium}
\altaffiltext{4}{UK ALMA Regional Centre Node, Jodrell Bank Centre for Astrophysics, School of Physics and Astronomy, University of Manchester, Oxford Road, Manchester M13 9PL, UK}
\altaffiltext{5}{Department of Physics and Astrophysics, Vrije Universiteit Brussel, Pleinlaan 2, 1050 Brussels, Belgium}
\altaffiltext{6}{Space Telescope Science Institute, 3700 San Martin Drive, Baltimore, MD 21218, USA}
\altaffiltext{7}{Astrophysics Group, Imperial College, Blackett Laboratory, Prince Consort Road, London SW7 2AZ, UK}
\altaffiltext{8}{Laboratoire AIM, CEA/DSM-CNRS-Universit\'e Paris Diderot, Irfu/Service, Paris, F-91190 Gif-sur-Yvette, France}

\altaffiltext{9}{Laboratoire d’Astrophysique de Marseille, UMR 6110 CNRS, 38 rue F. Joliot-Curie, 13388 Marseille, France}
\altaffiltext{10}{Department of Physics and Astronomy, University of California, Irvine, California 92697, USA}
\altaffiltext{11}{Istituto di Astrofisica e Planetologia Spaziali (INAF-IAPS), via del Fosso del Cavaliere 100, 00133, Roma}

\begin{abstract}  
	In this paper we present a catalogue of Giant Molecular Clouds (GMCs) in the Andromeda (M31) galaxy extracted from the {\it Hershel} Exploitation of Local Galaxy Andromeda (HELGA) dataset. GMCs are identified from the {\it Herschel} maps using a hierarchical source extraction algorithm. We present the results of this new catalogue and characterise the spatial distribution and spectral energy properties of its clouds based on the radial dust/gas properties found by Smith et al (2012). 236 GMCs in the mass range $10^4-10^7$\,M$_\odot$ are identified, their cumulative mass distribution is found to be proportional to $M^{-1.45}$ in agreement with earlier studies. The GMCs appear to follow the same cloud mass to $L_\mathrm{CO}$ correlation observed in the Milky Way. However, comparison between this catalogue and interferometry studies also shows that the GMCs are substructured below the {\it Herschel} resolution limit suggesting that we are observing associations of GMCs. Following Gordon et al. (2006), we study the spatial structure of M31 by splitting the observed structure into a set of spiral arms and offset rings. We fit radii of 10.5 and 15.5\,kpc to the two most prominent rings. We then fit a logarithmic spiral with a pitch angle of $8.9^\circ$ to the GMCs not associated with either ring. Lastly, we comment upon the effects of deprojection on our results and investigate the effect different models for M31's inclination will have upon the projection of an unperturbed spiral arm system. 
\end{abstract}



\section{Introduction}

\begin{figure*}
	\centering{
		\includegraphics[width=\textwidth]{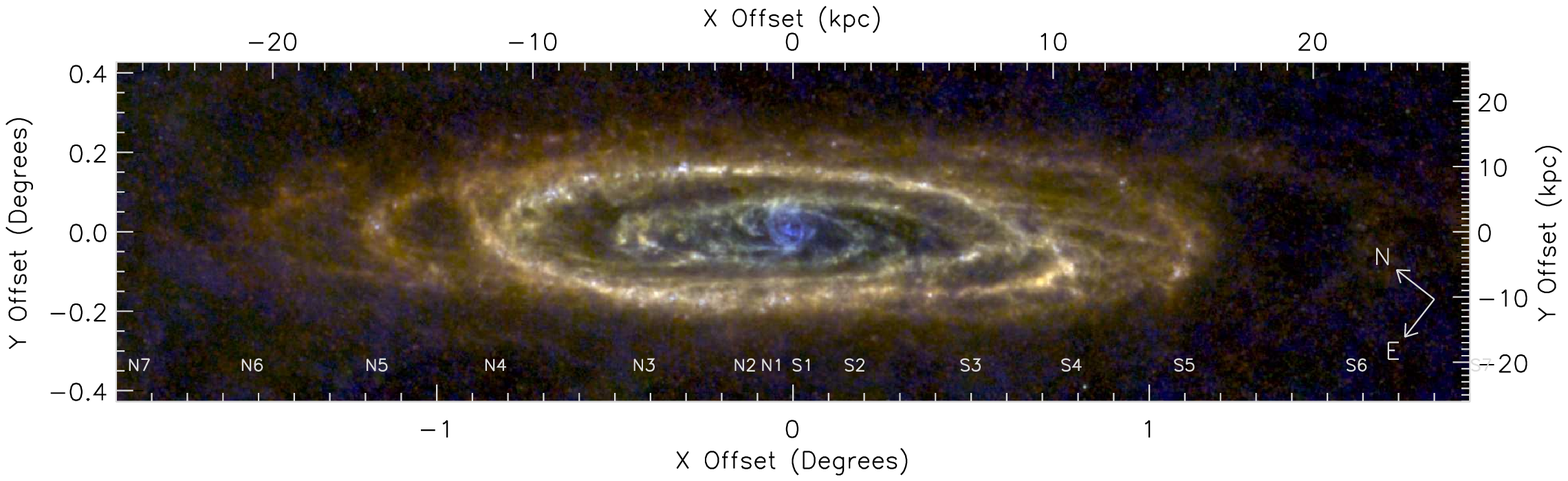} 
	}
	\caption{\label{fig:rgb} A three-colour image of M31 showing SPIRE 500$\mu$m (red) and 250$\mu$m (green) with PACS 100$\mu$m (blue). The map has been rotated by the assumed position angle of $38^\circ$. The position centre has an RA and Declination of $0^{h}\,42^{m}\,44\fs 330 41\degr\,16\arcmin\,07\farcs50$ (J2000), the direction of each axis is shown by the white arrows. The top and left axes show offsets calculated from the assumed distance to M31. The offsets of the spiral arm crossing points from \citet{1963esag.book.....B} are annotated as N1--6 and S1--6.   
	The green and blue data have been convolved to this study's working resolution (the 350$\mu$m beam FWHM, see text). The 500$\mu$m data is shown for comparison only and is left unconvolved. }
\end{figure*}

\begin{figure}
	\centering{
		\includegraphics[width=\columnwidth]{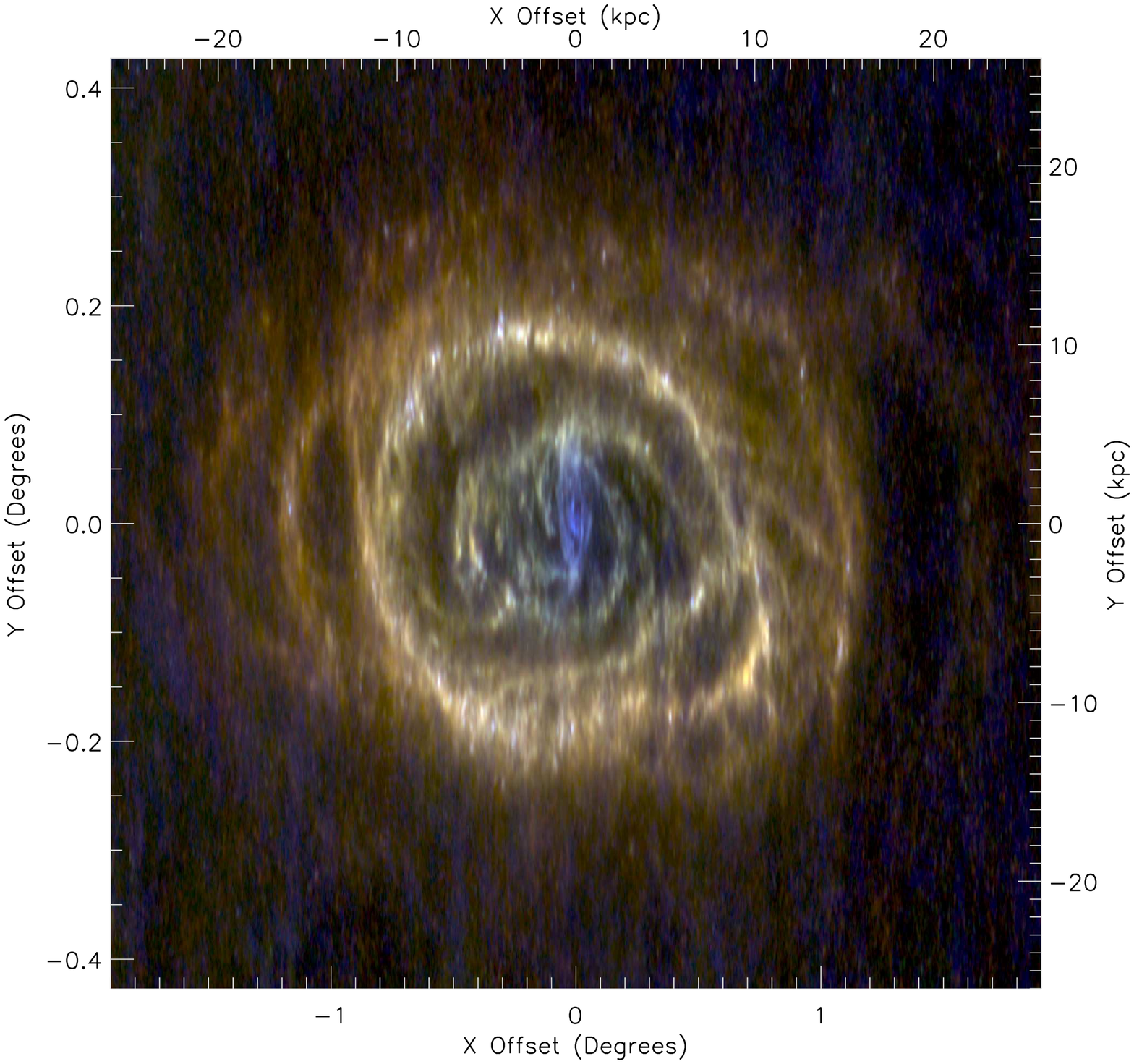} 
	}
	\caption{\label{fig:rgbproj}As Figure \ref{fig:rgb}, except that the y-axis has been deprojected using the assumed inclination angle of $77^\circ$. }
\end{figure}

The study of star formation within our own Galaxy is limited by our ability to resolve molecular clouds from the tangled web of the Galactic disc. The disc of the Milky Way has been mapped as part of the CO survey of \citet{2001ApJ...547..792D} and the {\it Spitzer} GLIMPSE \citep{2009PASP..121..213C} and MIPSGAL surveys \citep{2009PASP..121...76C}, amongst others, most recently by the Hi-GAL {\it Herschel} Open Time Key Project \citep{2010PASP..122..314M}. Nevertheless, ensemble studies of giant molecular clouds (GMCs) within our own Galaxy are still hampered by distance ambiguities and sampling limitations imposed by our own position within the Galactic disc. 

The solution to this is to study the same molecular clouds in nearby galaxies where we can observe the entire disc. The nearest spiral galaxy \citep[$785\pm25$\,kpc,][]{2005mcconnachie} to our own Milky Way is the Andromeda Galaxy (M31). It is the largest member of the Local Group of galaxies, of which our own Milky Way is the second largest. {\it Herschel} Exploitation of Local Galaxy Andromeda \citep[HELGA;][hereafter Paper I]{2011arXiv1112.3348F} is the first comprehensive, high-resolution far-infrared and sub-mm survey of M31 and its surroundings. There is a long established link between molecular gas emission and observations of dust \citep[e.g. ][]{1982AJ.....87.1165B}, so the {\it Herschel} data set allows us to trace emission from the GMCs in M31.

GMCs were first catalogued in the Milky Way as dark nebulae, dust silhouettes seen against a bright background star field \citep{1919ApJ....49....1B,1962ApJS....7....1L}, but a comparable catalogue of 730 dark nebulae in M31 was not published until the 1980s \citep{1980AJ.....85..376H}. Single dish CO observations mapped the molecular gas associated with these dark nebulae \citep{1981A&A....93L...1B,1988ApJ...328..143L}, but it was not until the first small interferometer maps were produced \citep{1987ApJ...321L.145V, 1993ApJ...406..477W,1995ApJ...444..157A,1998ApJ...499..227L} that individual GMCs could be resolved. Later studies were able to cover larger areas as \citet[][hereafter S08]{2000immm.proc...37S, 2008ApJ...675..330S} mapped 6 clouds and \citet[][hereafter R07]{2007ApJ...654..240R} mapped 67 clouds.  \citetalias{2007ApJ...654..240R} was the first study to perform a statistical analysis of GMCs in M31. Many of these studies (e.g., \citealt{1987ApJ...321L.145V}, \citetalias{2007ApJ...654..240R}, \citetalias{2008ApJ...675..330S}) have noted that, individually or in assemblage, M31's GMCs resemble those in the Milky Way. The HELGA data now allows us to determine whether this correspondence extends to the entire disc of M31.  

The dominant feature of M31 is a 100\arcmin\ diameter ring which appears in continuum emission from the infrared \citep{1984ApJ...278L..59H,1998A&A...338L..33H,2006ApJ...638L..87G} to the radio \citep{1984A&AS...56..245B}. The Ring is also detected in H-$\alpha$ \citep{1964ApJ...139.1045A,1994AJ....108.1667D} and carbon monoxide \citep{2006A&A...453..459N}. At the usual distance estimates to Andromeda, the Ring has a radius of 10\,kpc. \citet{2006ApJ...638L..87G} used deprojected {\it Spitzer} 24-$\mu$m data to fit a circle of radius 9.8\,kpc to the Ring and showed that its centre was offset from the centre of the spiral arm pattern. In \citetalias{2011arXiv1112.3348F} we showed how a comparison of {\it Herschel} and H{\sc i} atomic data revealed the presence of several low-intensity extended features we named E, F, and G at radii of 21, 26, and 31\ kpc, respectively. These features appear to form additional rings or arms beyond the well known 10\,kpc ring and the fainter 15\,kpc ring \citep{1998A&A...338L..33H,2006ApJ...638L..87G}.   

A key factor in deriving parameters from FIR observations is a practical knowledge of the dust grain properties. In \citet[][hereafter Paper II]{2012arXiv1204.0785S} we compared the HELGA data to the molecular gas as traced by  carbon monoxide line maps in order to examine the effects of metallicity gradients on the dust-to-gas ratio across the M31 disc. It was found that the gas-to-dust ratio $r_\mathrm{gd}$ had an exponential dependence with radius of the form $\log r_\mathrm{gd} = 1.1 + 0.0496 R$ where $R$ is the galactocentric radius. \citetalias{2011arXiv1112.3348F} found that the dust emissivity index, $\beta$, was ${\sim1.9}$ in the 10\,kpc ring, in broad agreement with studies of local Milky Way clouds \citep[e.g.,][]{2011A&A...536A..25P}. However, \citetalias{2012arXiv1204.0785S} also demonstrated that $\beta$ varied globally from a high of ${\sim}2.5$ in the centre to a value of ${\sim}1.7$ at large radii.  

\citet[][hereafter Paper III]{2012ford} combined {\it Galex} FUV and  {\it Spitzer} 24$\mu$m datasets to make a star formation rate (SFR) map of M31 and found a global SFR of $0.25$\,M$_\odot$\,yr$^{-1}$. This rate is a quarter of that of the Milky Way \citep{2010ApJ...710L..11R,2012ApJ...752..146L}, despite their masses as inferred from the motion of their satellites being comparable \citep{2010MNRAS.406..264W,2009ApJ...700..137R}. \citetalias{2012ford} also showed that M31 was positioned below the scatter of `normal' spiral galaxies on the Kennicutt-Schmidt plot \citep{1959ApJ...129..243S,1998ApJ...498..541K} of mass versus SFR surface densities. One of the questions that arises from these studies is whether the differences in SFR between M31 and the Milky Way is due to a difference in the number of giant molecular clouds (GMCs, the sites of star formation) or whether it is due to fundamental difference in the individual GMCs' properties. 

In this paper we use the HELGA data to analyse the population of GMCs and GMC complexes in M31. In Section~\ref{obs} we briefly describe the HELGA data and Section~\ref{extract} we describe the source extraction technique using the \textsc{CSAR} (Conservative Source AlgoRithm) dendrogram algorithm \citep{2013MNRAS.tmp.1218K}. In Section~\ref{prop} we examine the properties of the extracted sources and compare them with the observations of clouds from the Milky Way. Then in Section~\ref{structure} we re-examine the structure of M31 based upon the positions of the {\it Herschel} sources. 


\section{Observations}
\label{obs}

M31 was observed on the 18-20th of December, 2010 and the 23rd of January, 2011 (Observation Days 584-586 and 620) using the parallel-mode of the SPIRE \citep{2010griffin} and PACS \citep{2010poglitsch} cameras aboard the {\it Herschel Space Observatory} \citep{2010pilbratt}. These data used PACS filters centred at wavelengths of 100-$\mu$m and 160-$\mu$m, with angular resolutions of 12.5\arcsec\ and 13.3\arcsec\ respectively (accounting for the scan speed of 60\arcsec/s), and SPIRE filters centred at wavelengths of 250~$\mu$m, 350~$\mu$m, and 500~$\mu$m, with angular resolutions of 18.2\arcsec, 24.5\arcsec\ and 36.0\arcsec\ respectively. The observation strategy and data reduction methods are described in detail in \citetalias{2011arXiv1112.3348F}.

Figure \ref{fig:rgb} shows a three-colour image of M31 using SPIRE 500$\mu$m (red), 250$\mu$m (green), and PACS 100$\mu$m (blue) in rotated coordinates.  We use a position centre with a Right Ascension and Declination of $00^{h}42^{m}44\fs330~41\degr16\arcmin07\farcs50$ \citep[][the 2MASS catalogue position]{2006strutskie}. In common with other HELGA studies, we assume a distance to Andromeda of $785\pm25$\,kpc \citep{2005mcconnachie} and global inclination and position angles of $77\degr$ and $38^\circ$ respectively. A discussion of the assumed angles is included in Appendix \ref{deprojection}. At this distance, the final PACS and SPIRE angular resolutions are equivalent to spatial resolutions of 48--137~pc. 
 
The dominant feature at the longer wavelengths is the 10\,kpc Ring. This Ring is seen in Figure~\ref{fig:rgb} as the yellow/white-loop traced by strong SPIRE emission. Also visible in Figure~\ref{fig:rgb} is the colour difference between the cooler ring and the bluer, warmer galaxy centre. 

Andromeda is highly inclined and displays a significant warp at larger radii \citep[e.g., ][]{2009ApJ...705.1395C, 2010A&A...511A..89C}. The radius of the disc viewed in the {\it Herschel} images (e.g, Figure \ref{fig:rgb}) is $\sim20$\,kpc (similar to the radius out to which we can extract sources, see Figure \ref{fig:cloudprop1}), approximately half that for which H{\sc i} models have been computed \citep[e.g., ][]{2009ApJ...705.1395C, 2010A&A...511A..89C}. Figure~\ref{fig:rgbproj} shows the same as Figure~\ref{fig:rgb} except with the y-axis deprojected using the assumed inclination angle of $77^\circ$. We discuss the magnitude of the discrepancy between the assumption of flat or warped geometry on the deprojected structure in Appendix \ref{deprojection}. 


\section{Catalogue}
\label{extract}

\begin{figure}
\includegraphics[width=\columnwidth]{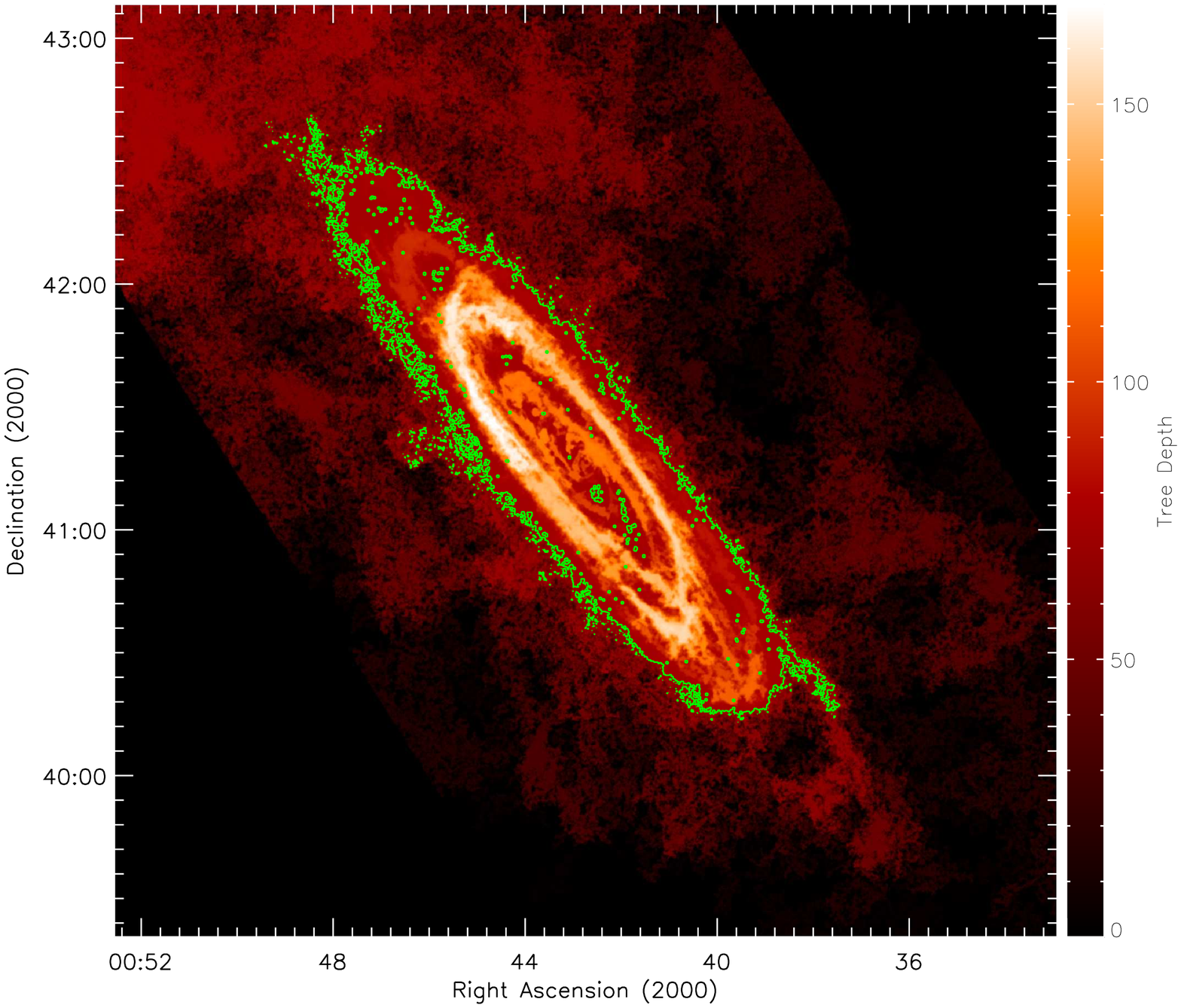}
\caption{\label{fig:depth} The dendrogram tree depth -- the number of branches between each and the base of the tree -- towards M31. The map has been smoothed slightly to enhance low-level features. The green contour shows the extent of the M31 tree branch.}
\end{figure}

\begin{figure*}
	\centering{\includegraphics[width=0.9\textwidth]{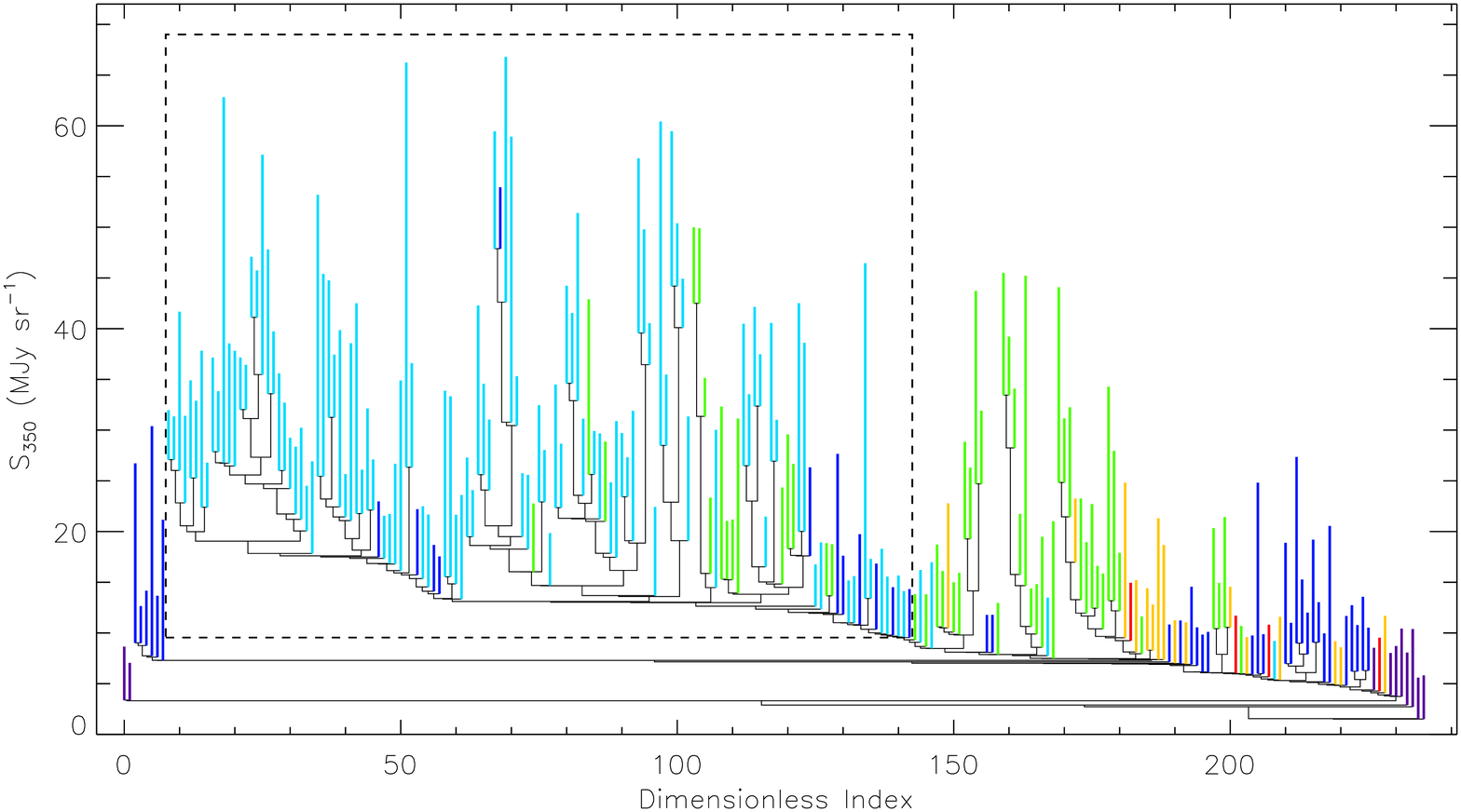}}
	\caption{\label{fig:dendro}A dendrogram of 350$\mu$m-intensity structure towards M31. The peak-to-background extent of each node is shown by the vertical bars. The coloured bars are the leaf-nodes, the colour denotes the galactocentric radius of each node (see Figure \ref{fig:dendromap}). The horizontal lines show the equivalent contour level at which two neighbouring nodes merge. The dashed box plotted over the dendrogram shows the branch containing the 10\,kpc Ring. }
\end{figure*}

\begin{figure*}
	\centering{\includegraphics[width=0.9\textwidth]{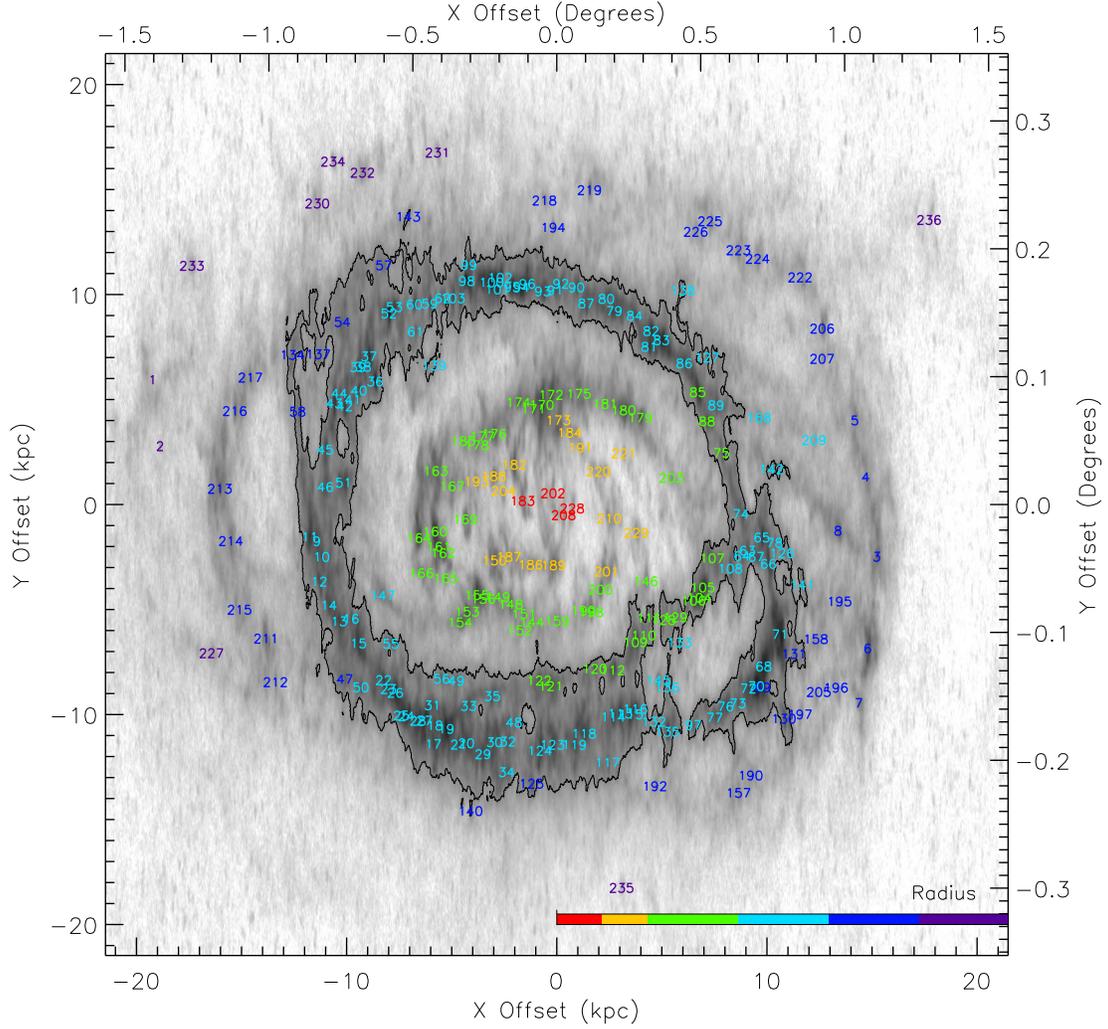}}
	\caption{\label{fig:dendromap} The relative positions of the leaf-node sources from Figure \ref{fig:dendro}. The greyscale shows SPIRE 350$\mu$m intensity. The contour shows the area enclosing the 10\,kpc Ring branch of the dendrogram. The axes show the offsets in kpc at the assumed distance to M31 and in degrees. The leaf-node labels are coloured depending on their galactocentric distance using the same colour scheme as \citetalias{2012ford}, the scheme is shown by the colour bar.  }
\end{figure*}

\clearpage

\subsection{Source Identification}

The emission from M31, as with any spiral galaxy, is highly hierarchical and is organised into tiers of rings, arms, complexes, and clouds. A source extraction algorithm must take this tree, the hierarchy of nested structures, into account. For this purpose we use the \textsc{csar} source extraction algorithm \citep{2013MNRAS.tmp.1218K}. \textsc{csar} works by processing the pixels in an image in order of descending flux, assigning each pixel in turn to a region centred on a local-maxima. A region is flagged as significant if it passes contrast (S/N $>5\sigma$) and size (larger than the telescope PSF) criteria. Neighbouring regions grow downwards in flux until they touch. If both regions are significant then a record of their state is made. The regions are then merged. The process continues until all pixels above a minimum value have been processed. 

\textsc{csar} is, in effect, walking through the binary structure tree of the map that it is being run upon. The tree is made from nodes, in effect single closed isophotal contours, and branches that relate how one node nests (encloses) another pair of nodes (contours). Nodes at the every top of the tree, the `leaf-nodes', contain no other nodes -- i.e., they contain no resolved substructure -- and are directly analogous to a normal source. At the base of the tree is the `root-node', this is the node which contains all the other nodes. Thus the tree describes an entire region as a set of closed isophotal contours and defines the region's structure by recording how those contours nest within each other.  

While the utility of using structure tree decomposition to study the hierarchical properties of molecular clouds has been known for some time \citep{1992ApJ...393..172H}, its practical application has only recently become routine \citep{2008ApJ...679.1338R, 2009Natur.457...63G, 2011arXiv1111.4391W}. The same theory has also been applied to the hierarchical relationship of stellar clusters in Local Group galaxies \citep{2010ApJ...725.1717G}. The theory behind the \citet{2008ApJ...679.1338R} and \citet{2011arXiv1111.4391W} codes and \textsc{csar} is similar, although \textsc{csar} is designed to work with monochromatic data.  

\textsc{csar} was run on the 350$\mu$m SPIRE image of M31. This was used as the working resolution/pixel grid as it improves upon the resolution of the 500$\mu$m data and still leaves at least three resolved data-points across the SED peak (enough points to fit a 2-parameter SED against). The extraction was performed on the data before it was deprojected in order to avoid problems introduced by a non-circular PSF. At each point on the map we define a tree-depth, this is the number of nodes that the flux from each point is shared between. Tree depth linearises the relative scaling between different levels in the tree and leaves only the statistically significant structure behind. Figure \ref{fig:depth} shows this for M31. Most of the structure in the 10\,kpc ring had upwards of 160 nodes/mergers between it and the root of the tree. Outside of M31 we see extended noisy structure that does not appear to be coherent, this is foreground Galactic cirrus. This cirrus is particularly visible in the north-east corner of the M31 field and is distinct from M31 in H{\sc i} line channel maps (see fig 4. of \citetalias{2011arXiv1112.3348F}). We exclude the cirrus from our analysis by `pruning' the tree back to the branch that we know just contains M31 emission. This is shown by the closed portion of the green contour on Figure \ref{fig:depth}.

\subsection{Dendrogram}

After pruning, the M31 tree was left with 471 nodes of which 236 were leaf-nodes. Figures \ref{fig:dendro} and \ref{fig:dendromap} show the \textsc{csar} results for the M31 branch of the 350$\mu$m structure tree. Figure \ref{fig:dendro} shows a dendrogram (a form of tree diagram used to represent structures in hierarchical datasets) of the extracted structure within M31. The vertical axis is the 350$\mu$m intensity, the horizontal-axis is an arbitrary-dimensionless index given to each leaf-node such as to unwrap the tree structure and clearly show the separate branches without them overlapping each other. The peak-to-background 350$\mu$m intensity of each node is shown by the vertical coloured bars. The horizontal lines show the equivalent contour level at which two nodes merge. Figure \ref{fig:dendromap} shows the positions of the extracted nodes plotted over a map of SPIRE 350$\mu$m intensity. The colour of the vertical bars in Figure \ref{fig:dendro} and the annotations in Figure \ref{fig:dendromap} are the same and show the distance of the leaf-nodes from the centre of M31. This is the same colour scheme as used by \citetalias{2012ford} and is shown by the bar in Figure \ref{fig:dendromap}.   

A striking feature of the dendrogram in Figure~\ref{fig:dendro} is that several large branches all merge at approximately the same intensity level (${\sim}13$\,MJy\,sr$^{-1}$), this is the 10\,kpc Ring. These branches are shown by the dashed box plotted over the dendrogram. The spatial extent of the region, the node just before the ring closes, is shown by the single contour in Figure \ref{fig:dendromap}.  We can now walk down the structure tree from the brightest source to the faintest. The structures that comprise the ring (the cyan markers) form several distinct complexes, but all merge together into a single structure at approximately the same intensity level (${\sim}13$\,MJy/sr). Several prominent structures interior to the 10\,kpc Ring (the green clouds) then merge with the tree before the exterior clouds (shown in blue and purple) connect. These exterior sources are distinct from background sources in that they are connected to M31 by contiguous emission. 

\subsection{Measured Properties}

\input{Andromeda_350res_table1_narrow_short}

The truncated tree contains 236 leaf-nodes.  We assume that these {\it Herschel} 350$\mu$m identified sources, which as stated have no resolved substructure, are GMCs or associations of several GMCs and are hereafter referred to simply as clouds. Table~\ref{tab:leafs} lists the properties of the clouds. The first column lists the catalogue number (a dimensionless index assigned during the plotting of the dendrogram, see below). All positions and sizes are calculated from the moments of each source's 350$\mu$m half-power contour. The second and third columns list the Right Ascension and Declination of the centroid of the half-power contour. The kiloparsec $X$ and $Y$ offsets in the rotated, deprojected frame ($\theta=38^\circ$, $i=77^\circ$) are listed in columns 4 and 5. Column 6 lists the Galactocentric Distance $R$ of the clouds. The positional accuracy is on the order of the the pixel size as the map is Nyquist sampled. At 350$\mu$m this is 8\,arcsec which equates to 30\,pc along the un-deprojected x-axis and $\sim140$\,kpc along the deprojected y-axis. Column 7 lists the geometric mean of the deconvolved FWHMs of each cloud.

The \textsc{csar} extraction produces a mask on the 350\,$\mu$m pixel grid for each cloud. Integrated flux densities are calculated by summing the pixels under each pixel mask at each wavelength. Before the fluxes were measured, the data were convolved to a common resolution (the SPIRE $350$\,$\mu$m PSF, 24\arcsec\ FWHM) using the \citet{2011PASP..123.1218A} convolution kernels and then co-aligned on the $350$\,$\mu$m pixel-grid (8\arcsec\ pixel width). We estimate the local pixel rms for each cloud by calculating the standard deviation of the pixels immediately adjoining its bounding-contour. The level of the isocontour is subtracted as a background from the pixels interior to it before the flux summation is preformed. Each cloud's spectrum is colour-corrected using the standard SPIRE \citep{2011bendo} and PACS \citep{2011muller} factors in an iterative loop. The median correction at each wavelength was $<2\%$. Most of the regions extracted by \textsc{csar} are larger than the telescope PSF (see Figure~\ref{fig:cloudprop1}) so we use the SPIRE extended source calibration.
  
{\it Herschel} colour-corrected fluxes at 350, 250, 160, and 100\,$\mu$m for each cloud are listed in columns 8 to 11 of Table~\ref{tab:leafs}. A 500\,$\mu$m flux is not listed as this data has a resolution lower than that of the extraction wavelength. One-sigma errors are listed for each source, these are the statistical errors based on the local pixel rms and do not include the systematic calibration uncertainties. Upper-limits are given for sources whose measured flux was less that $3\sigma$. Given that the majority of the sources are extended, the calibration error is 12\% for the SPIRE bands \citep{2011vaktchanov} and 10\% for the PACS \citep[][]{2012paladini}.

     
\section{Molecular Cloud Properties}
\label{prop}

\subsection{Cloud Size}

\begin{figure}
\centering{\includegraphics[width=\columnwidth]{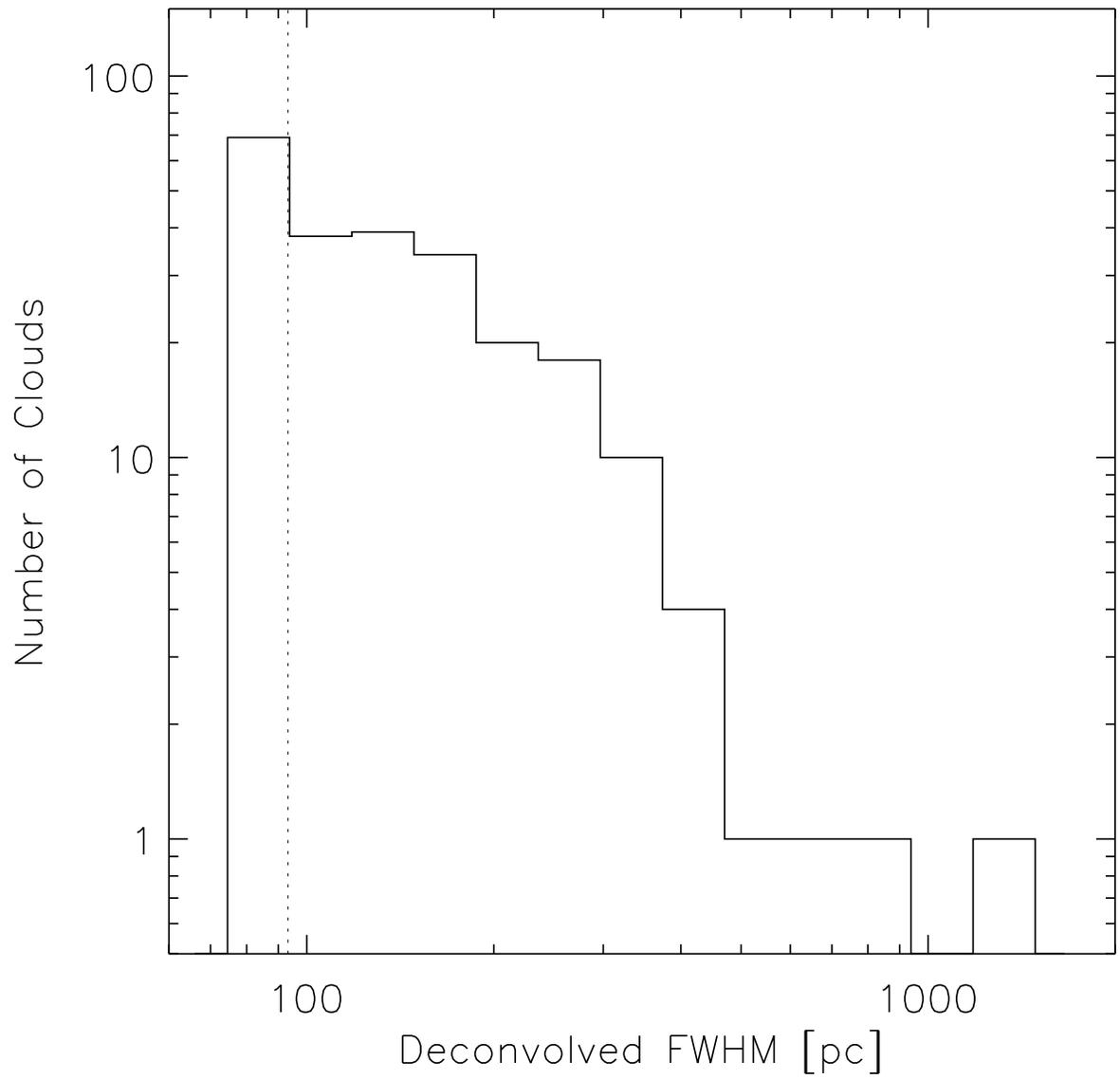}}
\caption{\label{fig:cloudprop1} Histogram of deconvolved source FWHMs. The vertical dashed line shows the beam FWHM. }
\end{figure}

For each cloud, a major and minor FWHM is calculated from the moments of the half-power contour. The geometric mean of the deconvolved major and minor FWHMs is then taken as the cloud's projected size as listed in column 7 of Table~\ref{tab:leafs}. The histogram of these sizes is shown Figure \ref{fig:cloudprop1}. The vertical dotted-line shows the 350\,$\mu$m beam FWHM, this is equivalent to 93\,pc at the assumed distance to M31. The bin to the left of this line is caused by the statistical scatter in the FWHM estimates for unresolved sources (all data points inside it have a {\it deconvolved} FWHM that is within 0.5 pixels of the resolution limit). It is assumed that the cross-section of the clouds are not effected by the projection of M31. 

The histogram shows a range of source sizes starting with unresolved sources. The majority of the clouds have sizes within 5 times the beam FWHM ($\sim500$\,pc). There is one source that is larger than 1\,kpc, this is located at X=-19, Y=4 which is a large, low-surface brightness arc. For comparison to these sizes, one of the nearest galactic GMCs, the Taurus molecular cloud, has a diameter of ${\sim}25$\,pc and the Gould Belt, the local system of GMCs, has a diameter of 1\,kpc. Additionally, the mean size of a GMC in the Milky Way is ${\sim}40$\,pc \citep{1979IAUS...84...35S} and in the LMC is ${\sim}30$\,pc \citep{2010MNRAS.406.2065H}. Thus, given this size distribution, the clouds we are extracting are probably complexes of GMCs and not the equivalent of individual GMCs. 

\begin{figure}
\centering{\includegraphics[width=\columnwidth]{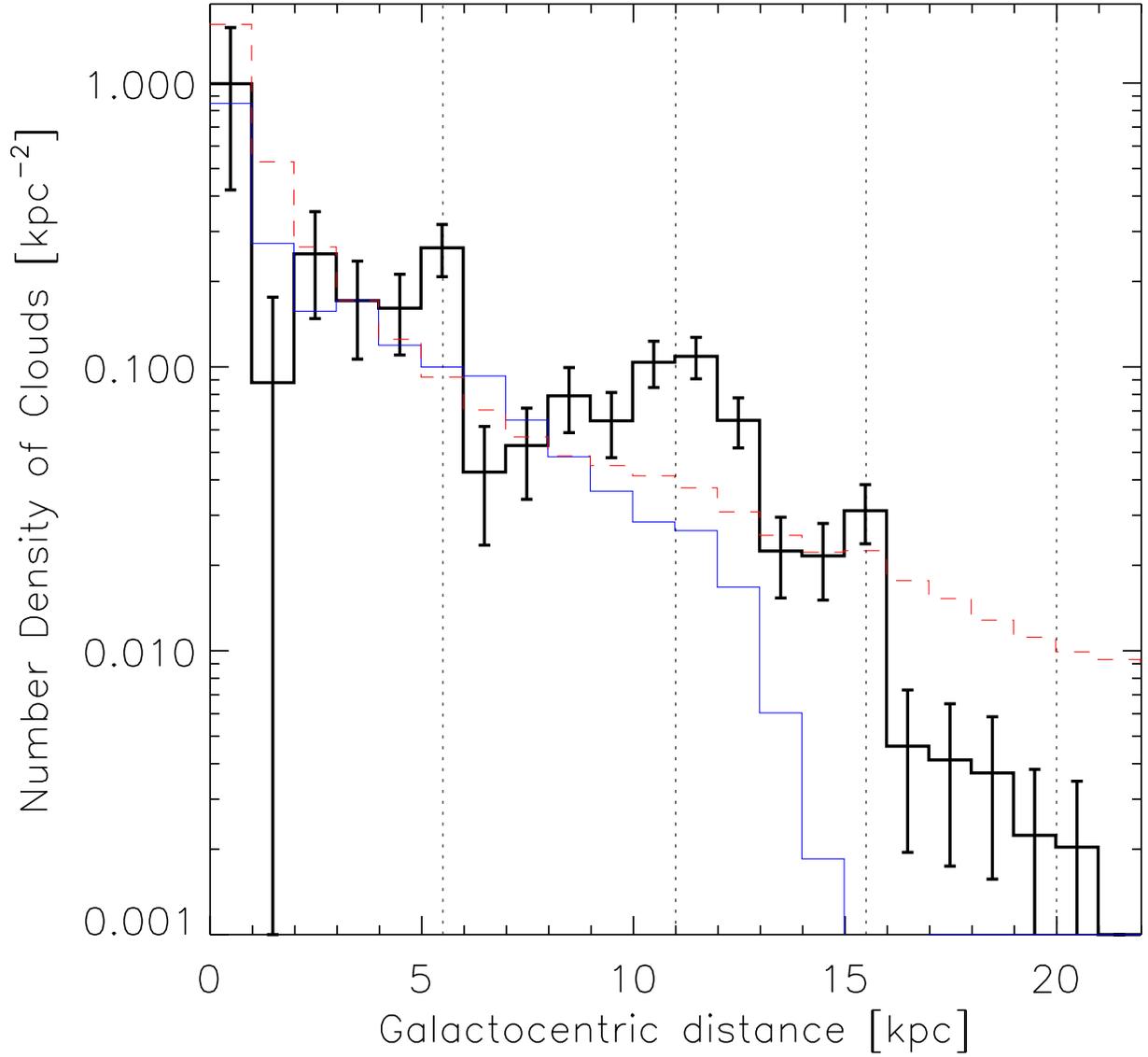}}
\caption{\label{fig:cloudprop2} The number density of sources with galactocentric distance. The vertical dashed line show the location of peaks coincident with ring structures. The solid blue curve shows the number density of dark nebulae from \citet{1980AJ.....85..376H} and the red dashed curve shows the surface brightness profile of M31 at 3.6$\mu$m. Both the red and blue curves have been normalised against the black curve at 3-4\,kpc. }
\end{figure}

Figure \ref{fig:cloudprop2} shows the number density of clouds (FIR sources) per square kiloparsec as a function of galactocentric distance. Error-bars are shown assuming normal errors. There is a clear downwards trend out to 20\,kpc, this is the region where the contiguous 350\,$\mu$m emission from Andromeda -- and thus the single-structure tree associated with it -- blends into the background. The fraction of optical light from the disc has also dropped off significantly at this radius \citep{2011ApJ...739...20C} although the disc features can be detected out to $40-50$\,kpc \citep{2005ApJ...634..287I,2011ApJ...739...20C,2011arXiv1112.3348F}. While there is scatter in this plot, it does show a series of peaks at ${\sim}5$\,kpc intervals (i.e., 5, 11, and 15\,kpc as shown by the dotted vertical lines) coincident with the observed rings at those distances. It is interesting to note that these follow the same pattern (${\sim}$20, 25, 30\,kpc) of FIR features in M31's outermost regions as detected by \citetalias{2011arXiv1112.3348F}. These features could be a system of weak resonant rings \citep[e.g., ][]{1996A&A...311..397J,1999Ap&SS.269...79B}. 

Also shown in Figure \ref{fig:cloudprop2} is the number density of dark nebula (solid blue curve) from \citet{1980AJ.....85..376H} and the surface brightness profile of M31 at 3.6$\mu$m (dashed red curve). Both the dark nebulae and 3.6$\mu$m profiles have been normalised against the distribution of FIR clouds at a radius of 3-4\,kpc. Both curves broadly follow the distribution of FIR clouds out to a radius of ${\sim}8$\,kpc with the exception of the peak at 5\,kpc. The 10\,kpc feature is seen as a significant enhancement above the 3.5\,$\mu$m profile, but it is not seen at all in the profile of dark nebulae. The distribution of dark nebulae drops off quite dramatically beyond 10\,kpc. It is possible that this is a selection effect, the distribution of the dark nebulae is correlated with the surface brightness of the disc because it is that which determines the background contrast and thus the chance of detecting a dark nebulae \citep{2008ApJ...680..349J}. Thus, it is possible that the optical surface brightness of M31 beyond 10\,kpc may not have provided sufficient contrast to discern nebula against in the \citet{1980AJ.....85..376H} survey. 

The peak of the size distribution of the \citet{1980AJ.....85..376H} dark nebulae is at ${\sim}100-150$\,pc, not dissimilar to our resolution limit. A comparison of source positions between the \citet{1980AJ.....85..376H} dark nebulae and our FIR selected clouds shows that only about 4.7\% of FIR cloud positions are within 100\,pc of a dark nebulae reference position. This only increases to 17\% if the search separation is increased to 200\,pc. \citet{1980AJ.....85..376H} himself noted the poor correlation between the distribution of dark nebulae and the distribution of atomic hydrogen. It is possible that dark nebulae poorly match the FIR clouds because they are just surface features seen {\it against} the bright disc of M31 whereas the FIR clouds trace emission through the entire depth of the disc. A similar feature is seen in the distribution of infra-red dark clouds (IRDCs) in the plane of the Milky Way \citep{2008ApJ...680..349J}. It is possible therefore that analysing dark features (dark nebulae) in a disc may not give a true representation of the properties of clouds in that galaxy \citep{2012MNRAS.422.1071W}.

\subsection{Cloud Mass}


For each cloud that has been detected (S/N$>3\sigma$) at 3 or more wavelengths between 100--350$\mu$m we follow \citetalias{2012arXiv1204.0785S} and fit a modified-blackbody function of the form 
\begin{equation}
	S_\nu = \frac{B_\nu(T_d) \kappa_\nu M_d }{D^2}, \label{smith}
\end{equation}
where $S_\nu$ is the flux density at frequency $\nu$, $B_\nu(T_d)$ is the Planck Function for a blackbody with temperature $T_d$, $M_d$ is the dust mass, and $D$ is the distance to the source. The dust absorption coefficient, $\kappa_\nu$, was parameterized as a power-law with the form $\kappa_\nu \propto \nu^\beta$ where $\beta$ is the dust emissivity index. The dust absorption coefficient was scaled from a reference value of 0.192\,cm$^2$\,g$^{-1}$ at 350\,$\mu$m \citep{2003ARA&A..41..241D}. This value is the same as used for \citetalias{2011arXiv1112.3348F} and \citetalias{2012arXiv1204.0785S}. The uncertainty in $\kappa_v$ could be as large as a few and is ignored when quoting uncertainties on the mass estimates.
 
We convert the dust mass, $M_d$, into a total mass (i.e., gas and dust), $M_\mathrm{cloud}$, taking into account the metallicity gradient in M31 by using the radial dust-to-gas and dust emissivity index relationships from \citetalias{2012arXiv1204.0785S}. There are therefore only two free parameters, $T$ and $M_\mathrm{cloud}$, for each fit. The Levenberg-Marquardt least-squares minimisation package MPFIT \citep{2009ASPC..411..251M} was used for all fitting and the fitting was done in $\log \nu$ vs. $\log \nu F_\nu$ parameter space. Hereafter, quoted cloud masses refer to the total mass of dust and gas unless stated otherwise. 

\begin{figure*}
	\centering{
		\includegraphics[width=0.48\textwidth]{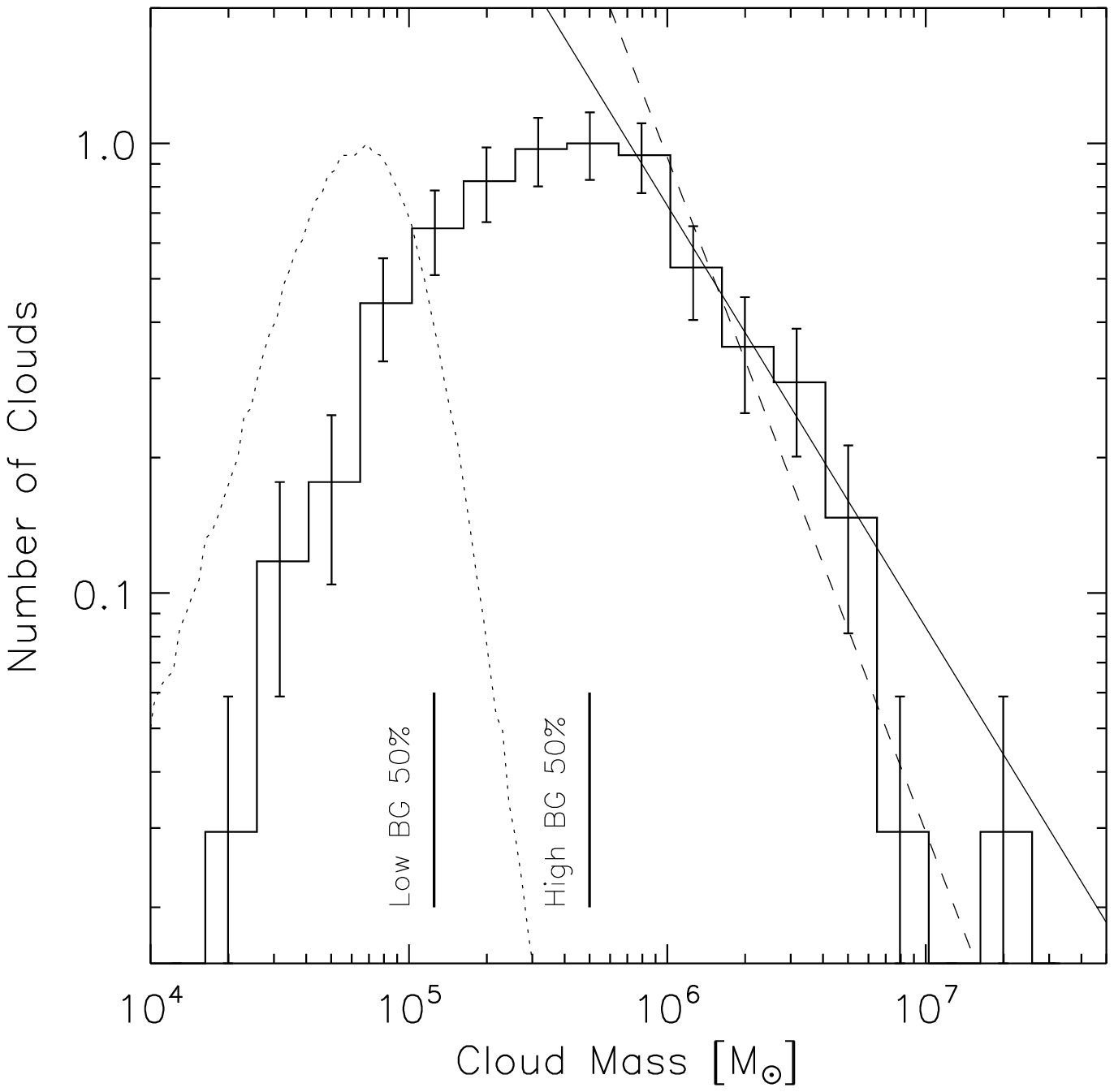}
		\includegraphics[width=0.48\textwidth]{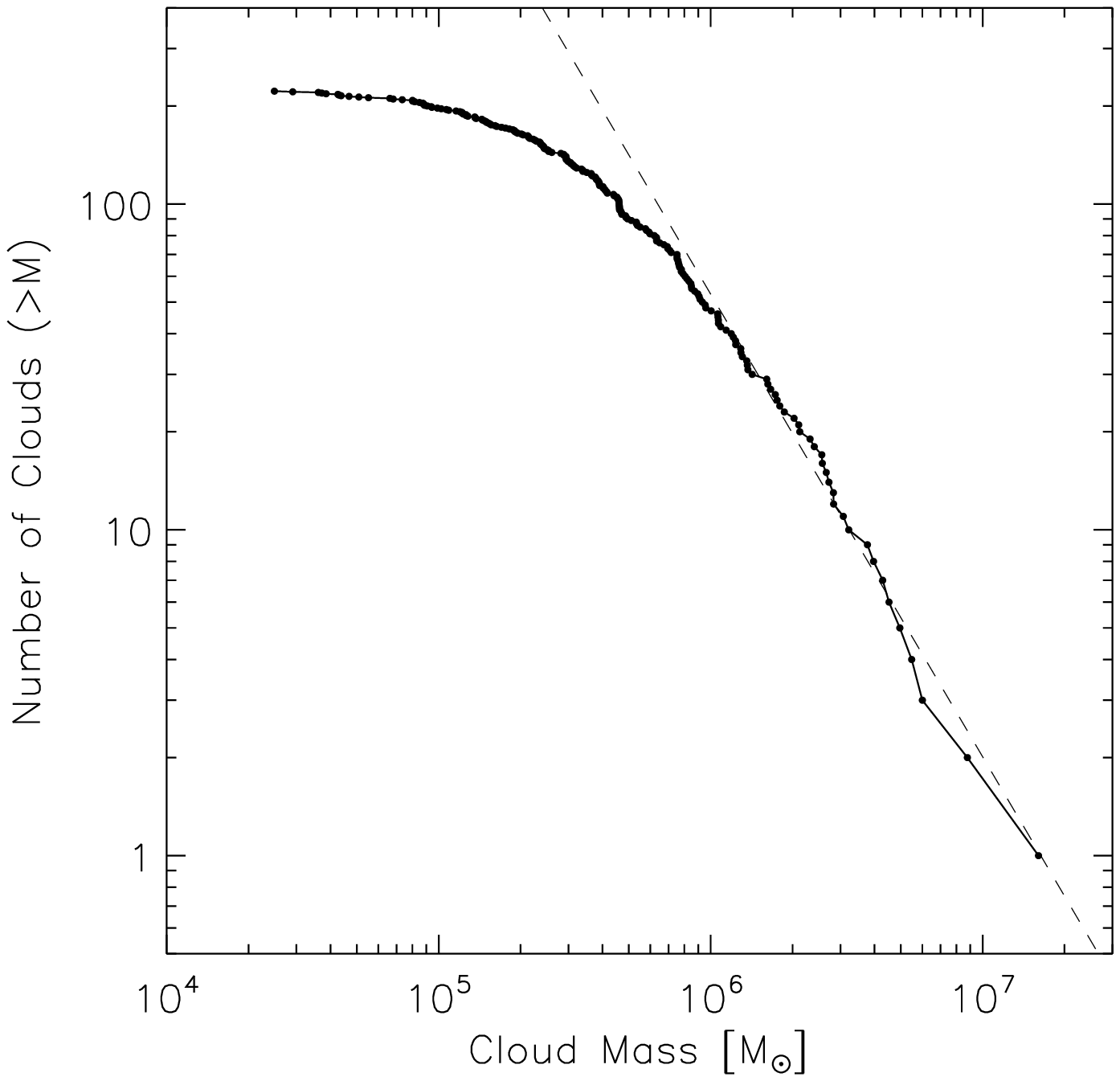}
	}
	\caption{\label{fig:mass} The mass distribution of clouds in M31. {\bf (left)} Histogram of $M_\mathrm{cloud}$. The solid powerlaw shows a best fit exponent of $\alpha_M = 0.94$, the dashed powerlaw show the equivalent $\alpha=1.5$ powerlaw for Milky Way clouds. The bars along the bottom edge show the 50\% point source completeness for low and highly structured background (see text for explanation). The dotted line shows our equivalent point source sensitivity once variations of dust temperature and properties have been accounted for. {\bf (right)} A cumulative histogram of total cloud mass for the M31 clouds. The dashed line shows a best fit powerlaw with an exponent of $p=-1.45\pm0.02$. } 
\end{figure*}

\input{Andromeda_350res_table2_narrow_short}

The fitted masses and temperatures are listed in columns 2 and 3 respectively of Table~\ref{tab:leafprops}. The clouds have a wide range of masses from $2.5\times10^{4}$ to $1.4\times10^{7}$\,M$_{\odot}$ with a median mass of $4.1\times10^{5}$\,M$_{\odot}$. Figure \ref{fig:mass} shows the normal $N(M_\mathrm{cloud}$) and cumulative $N(>M_\mathrm{cloud})$ mass distributions for the clouds. 

The higher-mass portion of the mass distribution can be fit by a powerlaw of the form $N(M) \propto M^{-\alpha_M}$. For the Andromeda clouds, we find a best fit of $\alpha_M = 0.94\pm0.2$ above a mass of $7.5\times10^5\,\mathrm{M}_\odot$. This is steeper than for clouds in the Milky Way which have a exponent of $\alpha_M{\sim}1.5$ \citep[e.g. ][]{1985ApJ...289..373S, 1987ApJ...319..730S, 2010ApJ...723..492R}. The best fit is shown by the solid line. A powerlaw with the same exponent as the Milky Way clouds is show for comparison. The Milky Way exponent is within three sigma of the Andromeda exponent and passes through most of the error bars above $10^6\,\mathrm{M}_\odot$. 
 
Assessing the completeness of the mass distribution of the extracted sources is complicated by the radial gradients in the properties of the dust and the non-uniform distribution of sources relative to those gradients. This is best illustrated by our point source mass sensitivity which is shown by the dotted curve plotted over the mass distribution. It was calculated by 100,000 repetitions of the mass calculation from our limiting flux sensitivity of $S_{350\mu\textrm{m}} = 55$\,mJy (a $5\sigma$ point source at 350$\mu$m, c.f. the \textsc{csar} extraction criteria) where the gas-to-dust ratio and the dust temperature have been sampled from normal distributions with the same mean and standard deviation as the sources in our catalogue ($r_{gd}=74\pm36$; $T=18\pm2$\,K). The resulting point source sensitivity is $6.6\pm4.5\times10^5\textrm{\,M}_\odot$. This corresponds to the lower tail of the mass distribution, but it does not account for the deviation of the distribution from a powerlaw between $10^5-10^6\,\mathrm{M}_\odot$. 

To simplify the assessment of our completeness we only consider point sources (the highest bin in the size histogram). Additionally, we consider two extremes of background -- a ``high'' $45\times12\arcmin$ background running along the SE portion of the 10\,kpc ring and a ``low'' background of the same size immediately to the SE of it.  We replicated the source extraction process on these backgrounds after injecting $350\mu$m point sources into the field with masses drawn from the fitted powerlaw distribution and with the same dust properties as those used in the previous test. A total of 10,000 sources were injected onto each background in batches of 25. It was found that that the mass distributions of synthetic point sources were 50\% complete above a mass of $1.25\times10^5\,\mathrm{M}_\odot$ for the low and $5\times10^5\,\mathrm{M}_\odot$ for the high backgrounds respectively. These limits are plotted on the mass distribution in Figure \ref{fig:mass}. Therefore, the departure of the mass distribution away from the power-law rise below $10^5$\,M$_\odot$ is due to incompleteness. 

The cumulative mass distribution for GMCs in a selection of nearby galaxies consists of a linear tail below ${\sim}10^{5}$\,M$_\odot$ that steepens at higher masses \citep{2007prpl.conf...81B,2010ARA&A..48..547F}. The cumulative mass distribution of M31's clouds is shown by the right-hand plot in Figure~\ref{fig:mass}. It is also flat below $10^5$\,$M_\odot$ and then begins to turn over between $10^5$--$10^6$\,M$_\odot$. A power-law of the form $N(>M_\mathrm{cloud})\propto M_\mathrm{cloud}^{p}$ was fit to the linear part of the distribution above a mass of $10^{6}$\,M$_\odot$. The value of the best-fit power-law exponent is $p=-1.45\pm0.02$. This is consistent with a value of $p=-1.55\pm0.2$ found from high-resolution CO mapping of a subset of M31 GMCs mapped with the BIMA interferometer (\citetalias{2007ApJ...654..240R},\citealt{2007prpl.conf...81B}). 

\subsection{Temperature and Luminosity}

\begin{figure*}
\centering{
\includegraphics[width=0.48\textwidth]{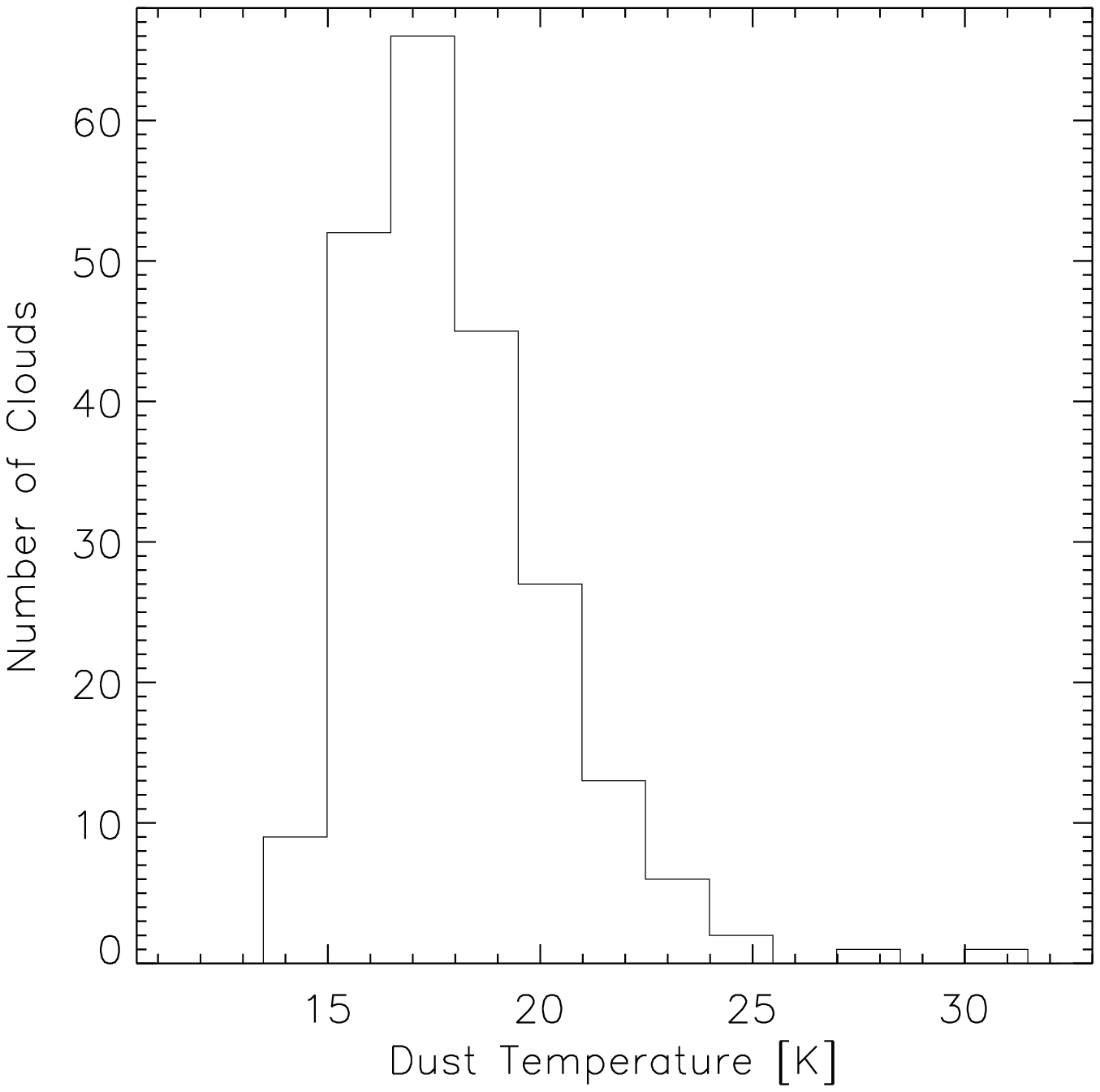}
\includegraphics[width=0.48\textwidth]{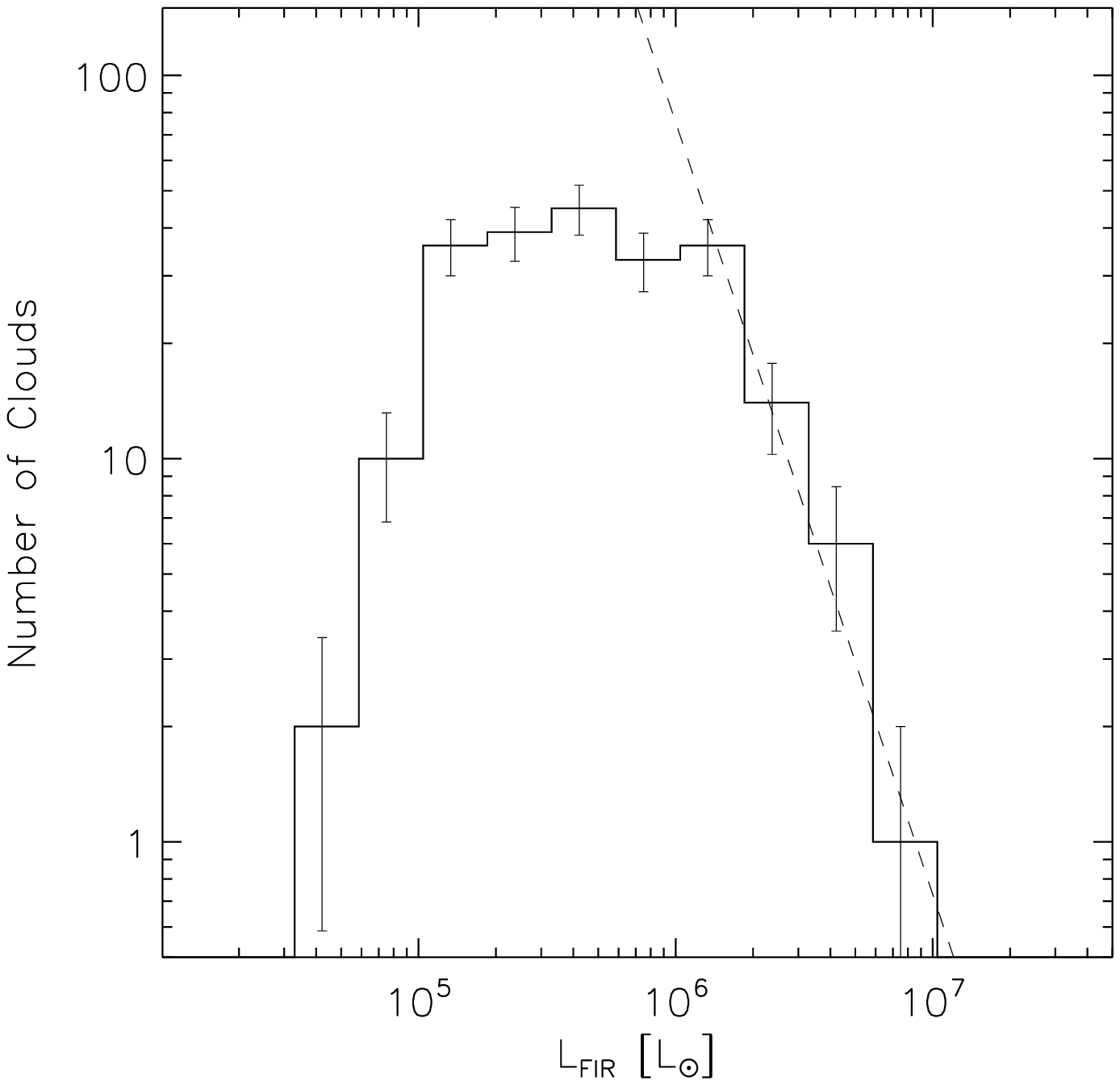} 
}
\caption{\label{fig:lumin} The temperature and luminosity of M31 clouds. {\bf (left)} Histogram of dust temperature fitted between the wavelengths of 100 and 350$\mu$m. {\bf (right)} Histogram of FIR luminosity $L_\mathrm{FIR}$ integrated beneath the best fit SED greybody. The dashed line shows a power-law fit with an exponent of $p=2.1\pm0.3$.  }
\end{figure*}

\begin{figure}
\centering{
\includegraphics[width=\columnwidth]{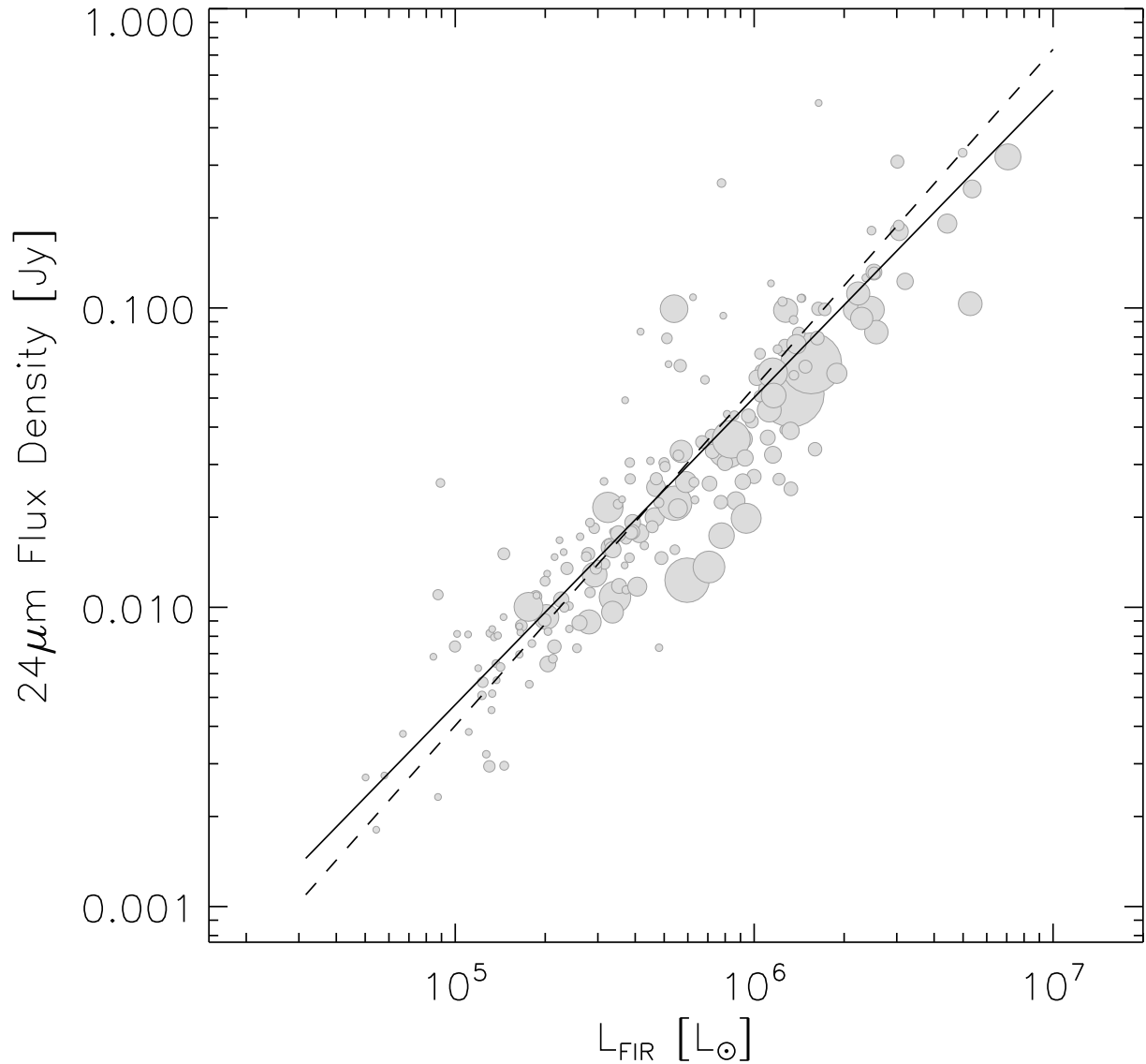}
}
\caption{\label{fig:24lumin}  
A plot showing the correlation between FIR luminosity and mid-infrared flux density. The width of the symbols shows the relative radius of the clouds. The solid line is a line of best fit, the correlation coefficient is 0.9.  The dashed-line indicates the slope of the relationship from \citet{2013arXiv1302.1858V}. 
}
\end{figure}
\clearpage

A histogram of temperatures obtained from the SED fitting is shown in the left panel of Figure \ref{fig:lumin}. The majority of cloud dust temperatures are in the range 15-25\,K with a median value of 18\,K. This is not unexpected as we are fitting a single temperature component over the wavelength range $100-350\mu$m and are therefore going to be dominated by the cold dust component. We can calculate the luminosity, $L_\mathrm{FIR}$, of this component by integrating beneath the fitted SED greybody in the range 10-1000$\mu$m. The individual values of $L_\mathrm{FIR}$ are listed in column 4 of Table~\ref{tab:leafprops} and their histogram is shown in the right-hand panel of Figure \ref{fig:lumin}. A power law fit to the histogram above $10^{6}$\,L$_\odot$ gives a best fit exponent of $p=2.1\pm0.3$. This is similar to the far-IR luminosity function for clouds in the Milky Way found by  \citet{1983MNRAS.203..955H} and to the CO luminosity function of clouds in M33 \citep{2012A&A...542A.108G,2007ApJ...661..830R}.  

Various continuum and multi-wavelength products have been used to calculate SFRs \citep[see ][ for reviews]{1998ARA&A..36..189K,2012ARA&A..50..531K} with many of these tracing their reasoning back to the Kennicutt-Schmidt Law \citep{1959ApJ...129..243S,1998ApJ...498..541K}. This Law assumes that the rate of star formation is proportional to some power of the interstellar gas surface density. Thus a measurement that determines the gas density or mass can be used as a proxy for the SFR. Two of these measures are the infrared luminosity and the mid-infrared flux density. Figure \ref{fig:24lumin} shows a plot of $L_\mathrm{FIR}$ with {\it Spitzer} $24\mu$m flux density for the M31 clouds. The {\it Spitzer} values were measured in the same manner as for the {\it Herschel} fluxes and used the same source masks.  About 11\% of the sources for which we have $L_\mathrm{FIR}$ estimates are undetected at 24$\mu$m and this ratio appears constant with Galactocentric distance. 

Figure \ref{fig:24lumin} shows that the 24$\mu$m flux density, which is a tracer of warm dust and thus a tracer for the amount of on-going star formation, correlates on a cloud/complex scale with the luminosity of cool dust, which traces the reservoir of gas available at the start of star formation process. Following \citet{2013arXiv1302.1858V}, the SFR calculated from the far-infrared continuum is $\mathrm{SFR}_\mathrm{FIR} \propto L_\mathrm{FIR}$ \citep{1998ARA&A..36..189K} while the star formation rate calculated from $S_{24\mu\mathrm{m}}$ is $\mathrm{SFR}_{24\mu\mathrm{m}} \propto S_{24\mu\mathrm{m}}^{0.88}$ \citep{2007ApJ...666..870C}. Eliminating the SFR between these gives $S_{24\mu\mathrm{m}} \propto L_\mathrm{FIR}^{1.13}$. This powerlaw is shown as the dashed line on Figure~\ref{fig:24lumin}. 

A best fit to the data gives,
\begin{equation}
	\log ( S_{24\mu\mathrm{m}} ) = (1.03\pm0.03 ) \log ( L_\mathrm{FIR} ) - \log ( 3.5\pm0.2 \times10^{-8} ) 
\end{equation}
where $S_{24\mu\mathrm{m}}$ is in Jy and $L_\mathrm{FIR}$ is in $L_\odot$. The best fit is shown by the solid line. The exponents of the relationship and best-fit are approximately equal. The correlation coefficient for this distribution is 0.90. \citet{2013arXiv1302.1858V} showed that the theoretical relationship also holds observationally for low-mass star formation regions within 1\,kpc of the Sun and for higher-mass star formation regions scattered through the Milky Way. The theoretical relationship appears to be consistent with the Andromeda data. This further reinforces the idea that the properties of the clouds in M31 are consistent with the expected properties of clouds in the Milky Way. The full SFR for these clouds and a detailed comparison with the results of \citetalias{2012ford} will be discussed in a follow-up paper.

\subsection{CO Luminosity}

\begin{figure}
	\centering{
		\includegraphics[width=\columnwidth]{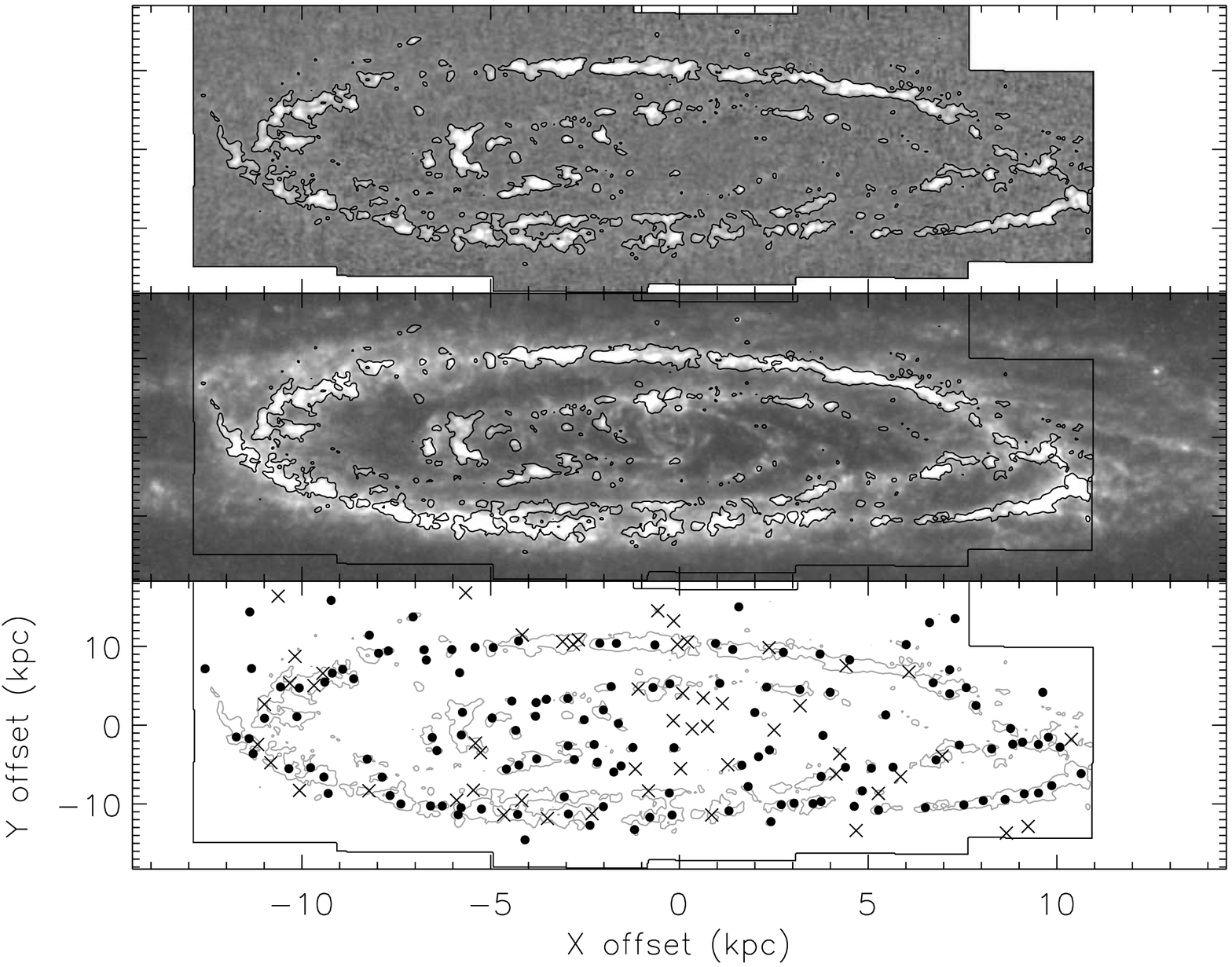}
	}
	\caption{Comparison of top: IRAM $^{12}$CO integrated emission \citep{2006A&A...453..459N} and middle: SPIRE 350$\mu$m dust emission towards the Andromeda Galaxy. The bottom panel shows which of the {\it Herschel} GMCs were detected (solid circle) or not-detected (crosses) in CO emission. The same 10-$\sigma$\ $^{12}$CO contour is plotted over each map. }
	\label{fig:comap}
\end{figure}

\begin{figure}
	\centering{
		\includegraphics[width=\columnwidth]{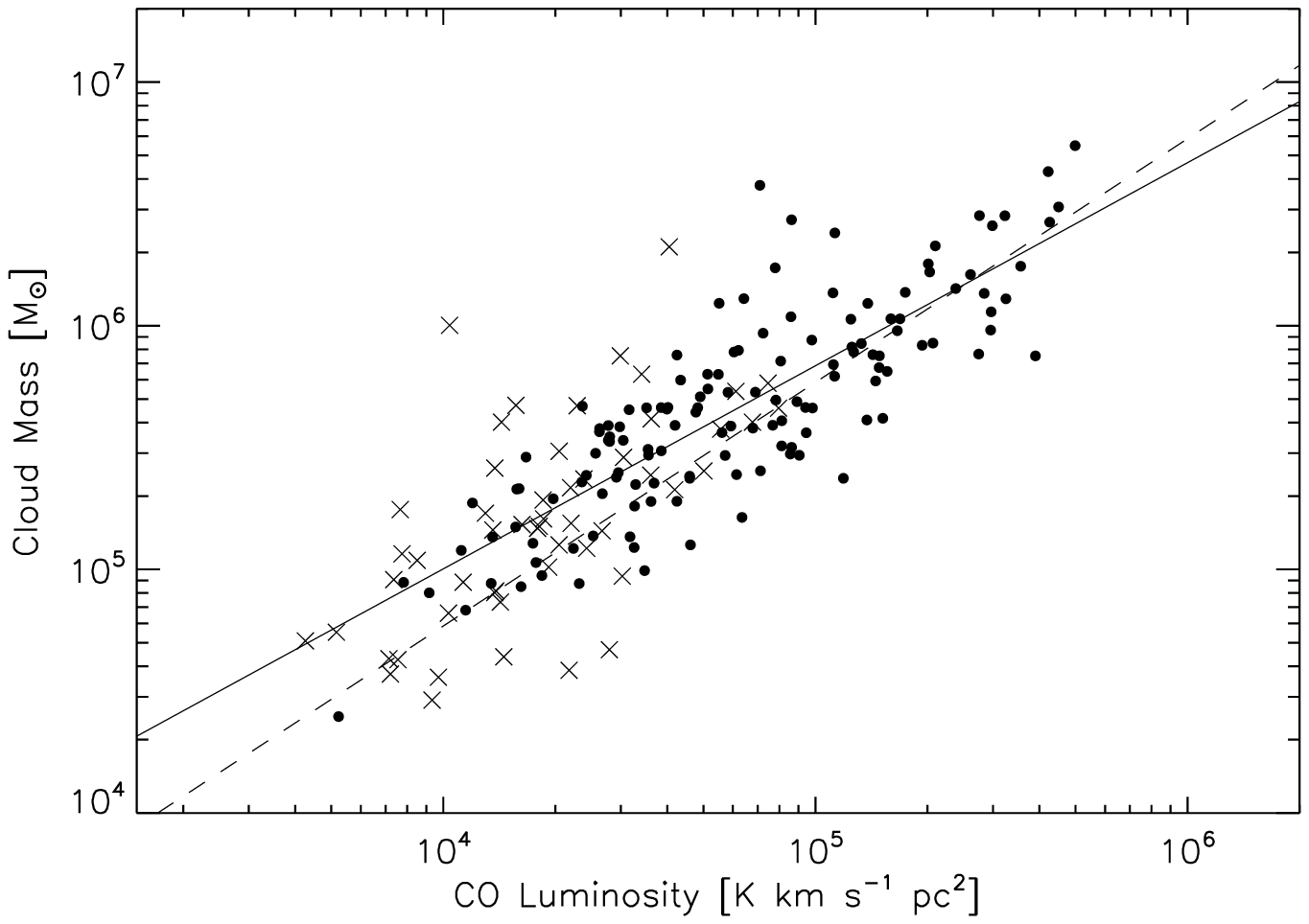}
	}
	\caption{Cloud $^{12}$CO luminosity versus total mass derived from the far-infrared. The solid line shows a best fit power with an exponent of $0.82\pm0.05$. The dashed line shows the $\alpha_{CO}$ relationship from \citetalias{2012arXiv1204.0785S}. }
	\label{fig:columin}
\end{figure}

The usual molecule for tracing giant molecular gas clouds is carbon monoxide, but there have been few comprehensive surveys across the entire disc of M31. \citet{2006A&A...453..459N} produced the first complete, sub-arcminute resolution CO study of M31, but there has not, as yet, been a published catalogue of individual CO clouds. We therefore use our catalogue to measure CO luminosities from the \citeauthor{2006A&A...453..459N} CO Map. 

\citet{2006A&A...453..459N} mapped M31 in the J=1-0 line of $^{12}$CO with a resolution of 23 arcsec using the IRAM 30-m telescope. This resolution is nearly equal to that of {\it Herschel} at 350$\mu$m, our working resolution, making comparison of our data with the Nieten CO relatively straight forward. Figure \ref{fig:comap} shows a comparison of the $^{12}$CO emission (top panel) and the 350$\mu$m emission (middle panel) towards Andromeda. The box outline shows the limit of the CO data. A single 10-$\sigma$\,$^{12}$CO contour is shown on both images where $\sigma=0.35$\,K\,km\,s$^{-1}$ \citep{2006A&A...453..459N}. The CO intensity traces well the peaks of the dust continuum. However, the extended dust component is not detected in the CO, showing that the CO emission is confined only to the densest regions.      

That dust emission traces CO luminosity well has been shown for a sample of galaxies including M31 \citep{2012arXiv1202.0547}, we can now test whether this holds for clouds inside M31 using the CSAR extraction contours to measure an integrated CO luminosity $L_\mathrm{CO}$ for each {\it Herschel} cloud using the Nieten CO map. CO luminosities in the range $10^4$ to $10^6$ K\,km\,s$^{-1}$\,pc$^{-2}$ were measured. The bottom panel of \ref{fig:comap} shows the location of the {\it Herschel} sources that were detected at greater than $3\sigma$ in the CO data (solid dots) and those that were not detected (crosses). 

Figure \ref{fig:columin} shows $L_\mathrm{CO}$ versus $M_\mathrm{cloud}$. The black dots show clouds that have a CO detection while the crosses show $3\sigma$ CO upper-limits for undetected clouds. There is a clear trend between the mass and CO luminosity. We test the correlation by preforming a linear regression to the CO detections. This gives a best fit of,
\begin{equation}
	\log(M_\mathrm{cloud}) = (0.82\pm0.05) \log(L_\mathrm{CO}) + (1.7\pm0.2)
\end{equation}
where $M_\mathrm{cloud}$ is in $M_\odot$ and $L_\mathrm{CO}$ is in K\,km\,s$^{-1}$\,pc$^2$. This is shown by the solid line. The correlation coefficient for this fit is 0.83 showing a reasonable correlation between the mass of a cloud derived from the {\it Herschel} data ($M_\mathrm{cloud}$) and that cloud's CO luminosity. The CO upper-limits appear to broadly follow the same trend, at least as far as low $L_{CO}$ correlates with low $M_\mathrm{cloud}$. 

The dashed-line on Figure \ref{fig:columin} shows the \citetalias{2012arXiv1204.0785S} relation of $\alpha_{CO} = M_\mathrm{cloud}(H_2)/L_\mathrm{CO} = 4.1$\,M$_{\odot}$\,pc$^{-2}$\,K$^{-1}$\,km$^{-1}$\,s\, under the assumption that $M_\mathrm{cloud}$ is  $70\%$ H$_2$ by mass \citep[proto-solar abundance, e.g.,][]{2009ARA&A..47..481A}.

\citet{1987ApJ...319..730S} presented a CO survey of clouds in the Milky Way observed with the 14-m FCRAO antenna. The implied sizes for their clouds are lower than we have measured for M31, but the range of CO luminosities are almost identical to those we calculate.  This supports the proposition that we are not resolving individual clouds and are instead resolving assemblages of individual clouds. The upper range for both surveys is $L_\mathrm{CO} = {\sim}10^6$\,K\,km\,s$^{-1}$\,pc$^{-2}$. They find a best fit between the virial mass $M_\mathrm{virial}$ of each cloud and $L_\mathrm{CO}$ of
\begin{equation}
	\log(M_\mathrm{virial}) = 0.81 \log(L_\mathrm{CO}) + 1.6
\end{equation}
This is virtually identical to the power-law that we fit to the clouds in Andromeda. 

\citeauthor{1987ApJ...319..730S} related the optically thick $^{12}$CO luminosity, which is proportional to the cloud's cross section, to the virial mass using a size-linewidth relation to give $M_\mathrm{virial} = 43 L_\mathrm{CO}^{4/5}$\,M$_\odot$. This relationship would hold for virialised clouds. That we obtain the same mass-luminosity relationship in M31 implies that these clouds are virialised on some level. This would probably not be on the complex scale, but could be at some spatial scale below our resolution limit (i.e., the complexes are made up unresolved virialised units).

\subsection{Comparison With Interferometry studies}

\begin{figure*}
\centering{
	\includegraphics[width=\textwidth]{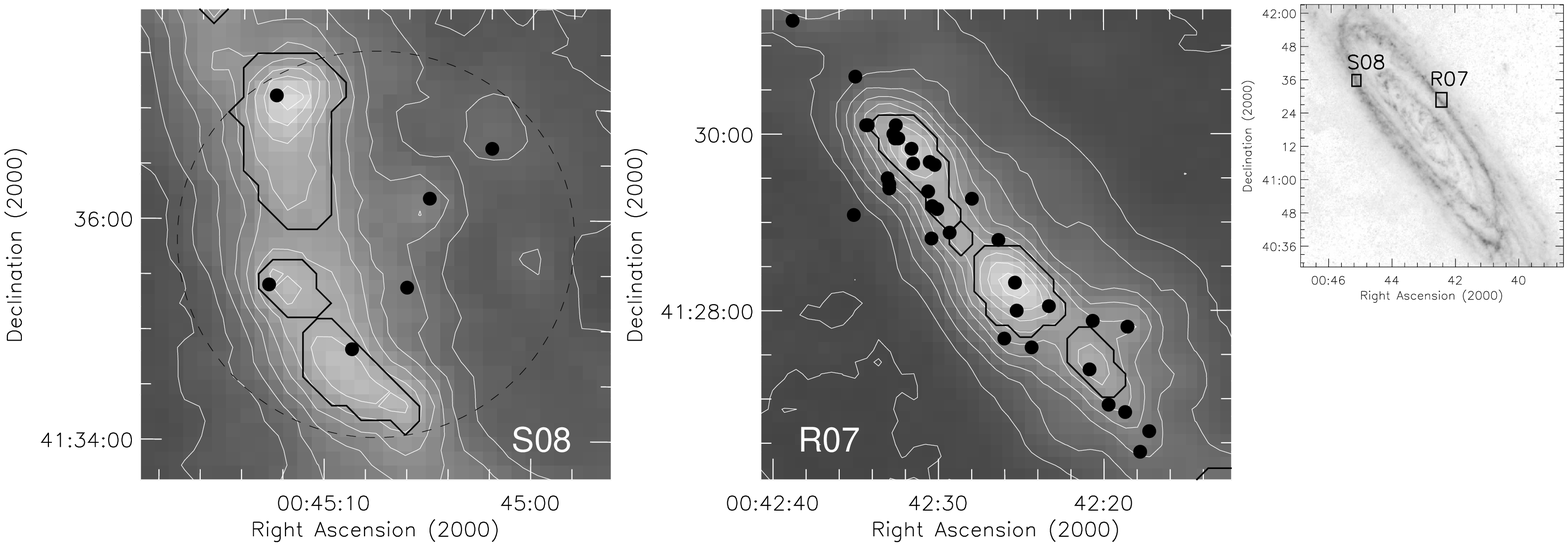}
}

\caption{\label{fig:sheth}Comparison of SPIRE 350$\mu$m results to BIMA interferometric sources. Left-panel based on Fig. 1 of \citet{2008ApJ...675..330S}. Middle-panel based on Fig. 3 Field A from \citet{2007ApJ...654..240R}. The dots show the positions of interferometry CO clouds. The large dashed circle shows the limit of the \citet{2008ApJ...675..330S} map. The thick-black contours show the location and extent of sources from this paper. The greyscale and contours are SPIRE 350$\mu$m data. The contours are spaced at 5$\sigma$ intervals. The smaller right-hand panel shows the location of the other two panels. }
\end{figure*}

Studies of individual giant molecular clouds in M31 with interferometers have been made \citep{1987ApJ...321L.145V, 1993ApJ...406..477W,1995ApJ...444..157A,1998ApJ...499..227L}, the increasing sensitivity of millimetre interferometers has meant that studies of more than a few clouds at a time is now possible (R07, S08). R07 mapped a $\sim7$\,kpc arc along a northwestern section of the 10\,kpc ring (among other fields). They detected 19 clouds which they could accurately resolve the properties of and a further 48 unresolved clouds. S08 mapped a single 2-arcmin diameter field across a northeastern section of 10\,kpc ring, detecting 6 clouds. 

Figure \ref{fig:sheth} shows a comparison of the results from these two studies using the BIMA interferometer and the {\it Herschel} data presented in this paper. The greyscale and contours are SPIRE 350$\mu$m dust emission with contour spacings of 5$\sigma$. The middle panel shows the region coincident with the R07 Field A while the left-hand panel shows the S08 field. The BIMA clouds are shown by the markers (only the resolved sources are shown for R07 field). R07 and S08 have resolutions that are $\sim3$ times the resolution of the 350$\mu$m {\it Herschel} maps. Comparison with the R07 field shows that the leaf nodes identified in this paper can break into multiple objects when viewed with an interferometer. Typically there is one BIMA source identified with the {\it Herschel} peak and several more sources clustered around it. The S08 field shows a similar pattern. They did not break down their clouds into sub-fragments as R07 did, but the detailed structure is still visible in their original maps. 

The comparison of the BIMA data to the {\it Herschel} regions shows that it is correct to think of the {\it Herschel} regions as complexes of giant molecular clouds and not individual clouds. This comparison also illustrates the difficulty of comprehensively mapping a source as large as M31 with sufficient resolution to resolve individual star formation regions. The speed and sensitivity of ALMA would go a long way to solving this, but its relatively high declination and large size makes observing M31 challenging -- if not virtually impossible -- from ALMA's location. It would, however, make an excellent target for the proposed NOEMA array at IRAM. 


\section{Global Structure}
\label{structure}

\begin{figure}
	\centering{
		\includegraphics[width=\columnwidth]{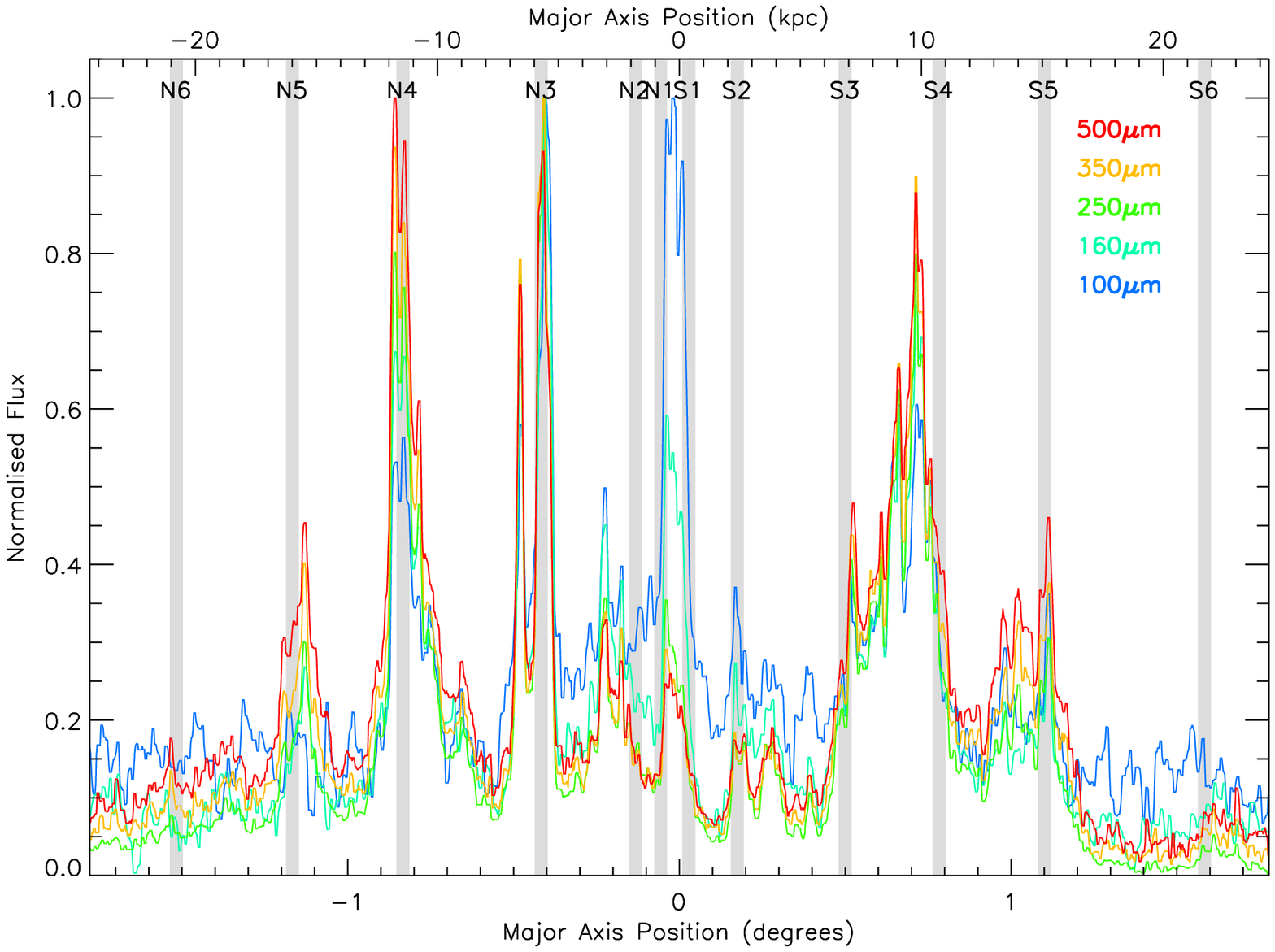}
	}
	\caption{\label{fig:slice}Normalised intensity slices along the major-axis of M31 from PACS 100$\mu$m (blue) to {\it Herschel} SPIRE 500$\mu$m (red), the key in the top-right shows the colour assigned to each wavelength. The positions of the arm crossing points from \citet{1963esag.book.....B} are annotated as N1--6 and S1--6.}
\end{figure}

Figure \ref{fig:slice} shows a series of intensity slices taken at different wavelengths along the major-axis of M31. The data is normalised against the peak flux in each band and is convolved to a resolution of 24\arcsec\ FWHM (the same as was used for the flux density measurements). The 500$\mu$m data has been left unconvolved as its PSF is larger than the 350\,$\mu$m PSF. The annotations show the positions of the arm crossing regions described by \citet{1963esag.book.....B} from his survey of M31 using the Mount Wilson 100-inch telescope. These are labelled numerically proceeding from the centre outwards in northerly and southerly directions. The innermost N1 and S1 arms, and to an extent the N2 and S2 crossing points, show an excess of emission at shorter wavelengths. This is the region that appears blue in the false-colour image in Figure \ref{fig:rgb} indicating the presence of hot dust. The N3 crossing point shows a strong peak of emission at long wavelengths relative to the other inner arms.

\citetalias{2011arXiv1112.3348F} reported the existence of a series of low brightness rings and structures surrounding M31 in the {\it Herschel} maps. \citetalias{2011arXiv1112.3348F} confirmed the detection of a 15\,kpc ring previously seen with {\it Infrared Space Observatory} \citep{1998A&A...338L..33H} and {\it Spitzer} \citep{2006ApJ...638L..87G}. The 15\,kpc ring (equivalent to ${\sim}1^\circ$ at 785\,kpc) is seen in Figure \ref{fig:slice} as the peaks coincident with Baade's N5 and S5 arm crossing points. In addition to the 15\,kpc ring, \citetalias{2011arXiv1112.3348F} reported three additional structures they labelled E, F and G at major-axis distances of ${\sim}21$, ${\sim}26$, and ${\sim}31$\,kpc (equivalent to ${\sim}1.5^\circ$, ${\sim}1.9^\circ$, and ${\sim}2.25^\circ$ at 785\,kpc). The E feature is visible as the faint red band on the right of Figure \ref{fig:rgb} and a minor rise in emission associated with the S6 arm crossing in Figure \ref{fig:slice}. The S7 crossing point is not shown, but at $1.9^\circ$ from the centre, it would be coincident with the F feature in the {\it Herschel} maps.   

Analysis of M31's spiral structure is hampered by the heavy disruption to the galaxy in the southern quadrant. The most significant feature is a $30^\circ$ wide break in the ring at a position of (8, -8) kpc coincident with the position of star forming cloud NGC 206 (visible in Figure~\ref{fig:anterior}). There is further evidence of this disruption in the arm crossing slice shown in Figure \ref{fig:slice}. The northern arm segments are all coincident with the strongest emission peaks. However, the southern crossing points only show a weak coincidence, if any, with the strongest emission peaks. There is some emission peaking with S2 and S3, but the brightest peaks actually occur between the S3 and S4 as part of an extended plateau of emission which stretches from S2 to S5. The outer most of these two peaks, at ${\sim}40$\,\arcmin, is coincident with the sweep of the Ring and the possible spiral arm pattern. 

In order to more accurately describe the properties of the Andromeda spiral arms we follow \citet{2006ApJ...638L..87G} and analyse the 10\,kpc Ring separate from the arms themselves. \citet{2006ApJ...638L..87G} described the structure of M31 with a classic two-arm logarithmic spiral and an offset ring. In addition to this radial profile analysis we attempt a fit to the 15\,kpc ring. A summary of the results is given in Table~\ref{tab:struct}.  

\subsection{The 10\,kpc Ring}

\begin{figure*}
\centering{
	\includegraphics[width=0.48\textwidth]{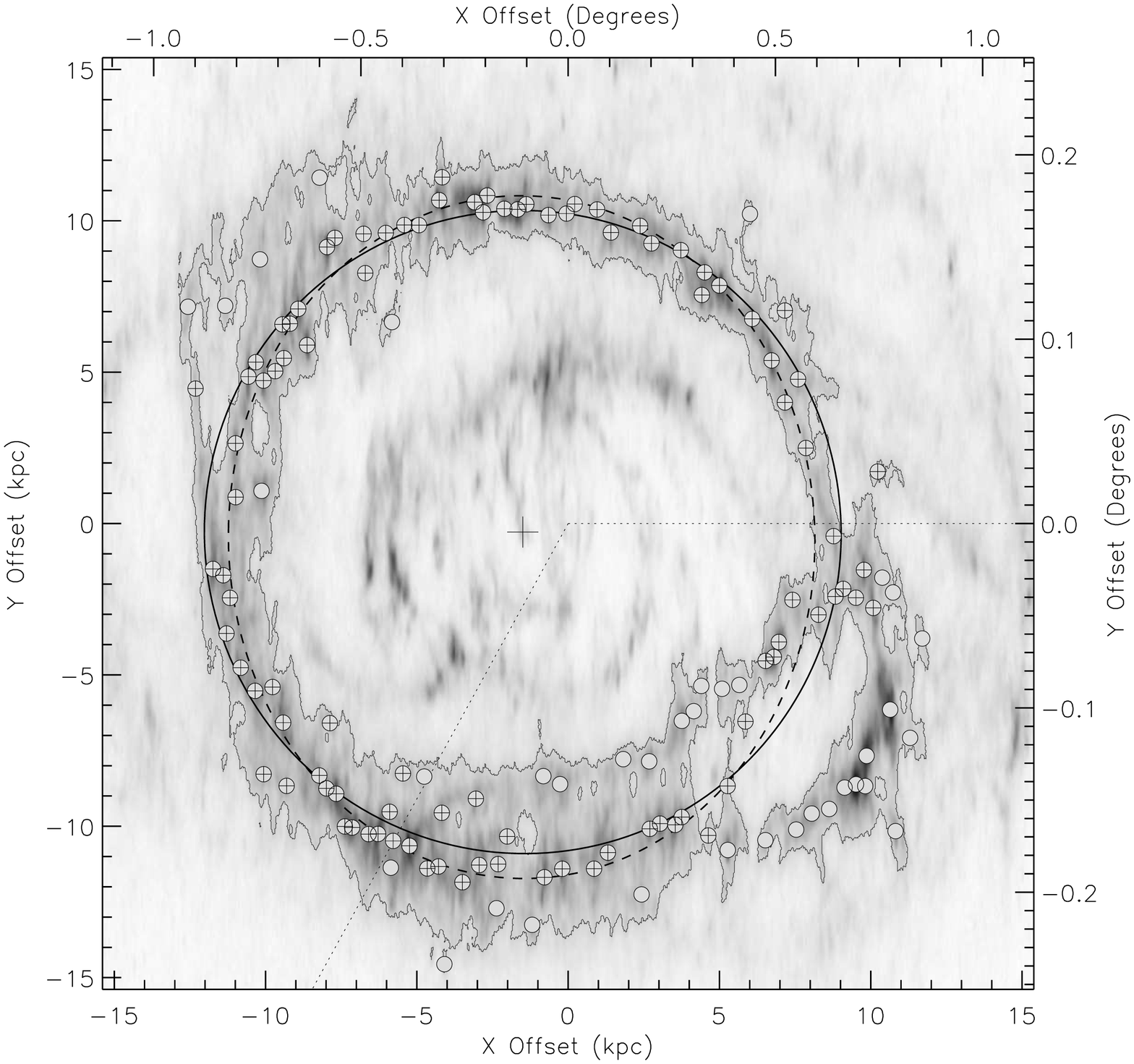}
	\includegraphics[width=0.45\textwidth]{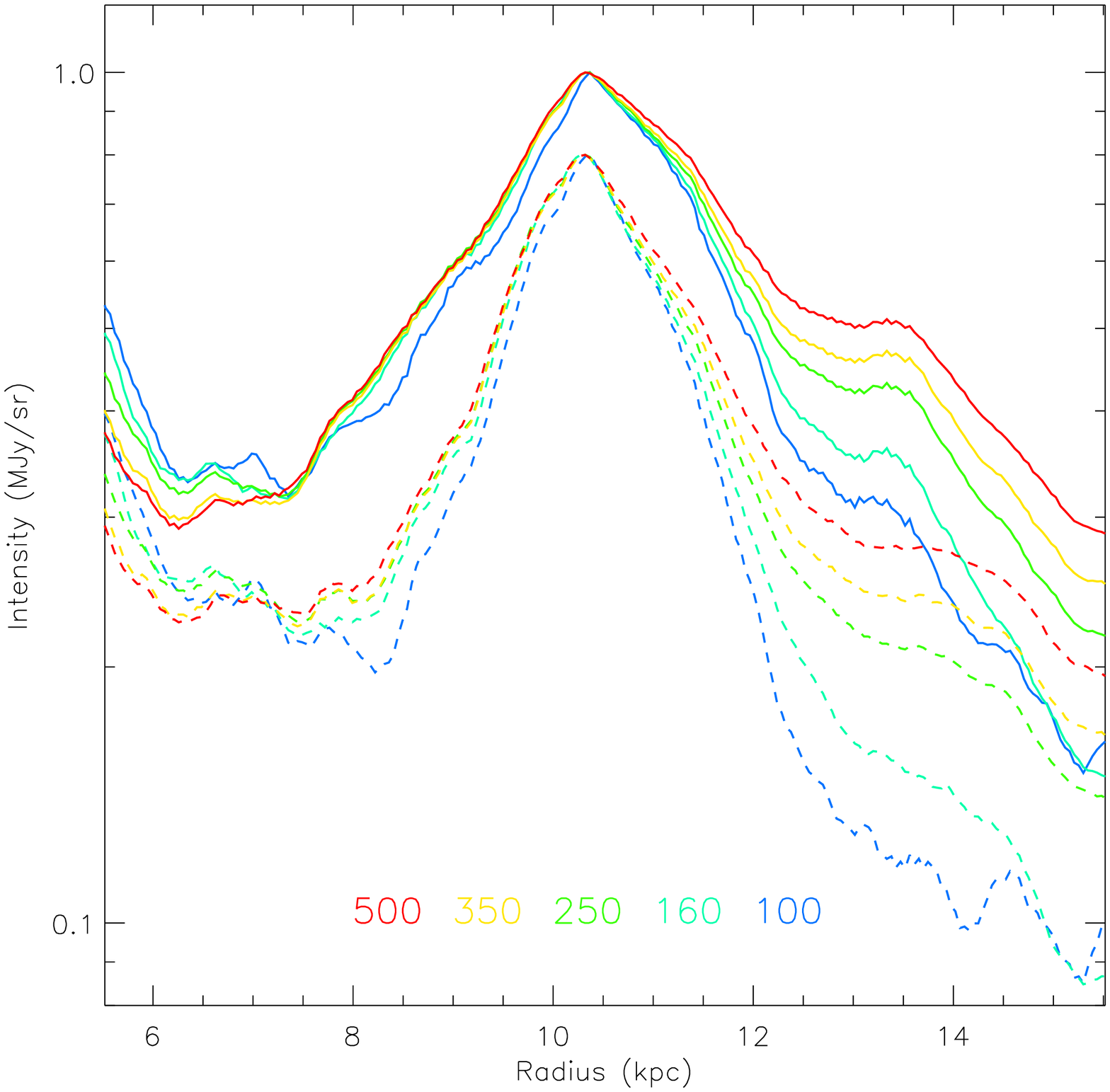}
}

\caption{\label{fig:anterior}
	{\bf (left)} An enlargement of the centre of Andromeda. The greyscale is the deprojected SPIRE 350$\mu$m map. The black contour is the 10\,kpc Ring node from Figure~\ref{fig:dendro}. The clouds of the Ring are shown by the circle markers, those involved in the final Ring fit have a cross inside the markers (see text for details). The solid and dashed circles show the Ring fitted with assumed inclination angles of 77$^\circ$ and 75$^\circ$ respectively. The dotted lines show the PA range 120-240$^\circ$. 
	{\bf (right)} A circularly-averaged profile of 10\,kpc Ring. The solid lines show the {\it Herschel} wavelengths (see key for colours) averaged over all position angles. The dashed lines show the profiles excluding the data from the position angle range 120-240$^\circ$.   
	 }
\end{figure*}

\begin{figure}
	\centering{
		\includegraphics[width=\columnwidth]{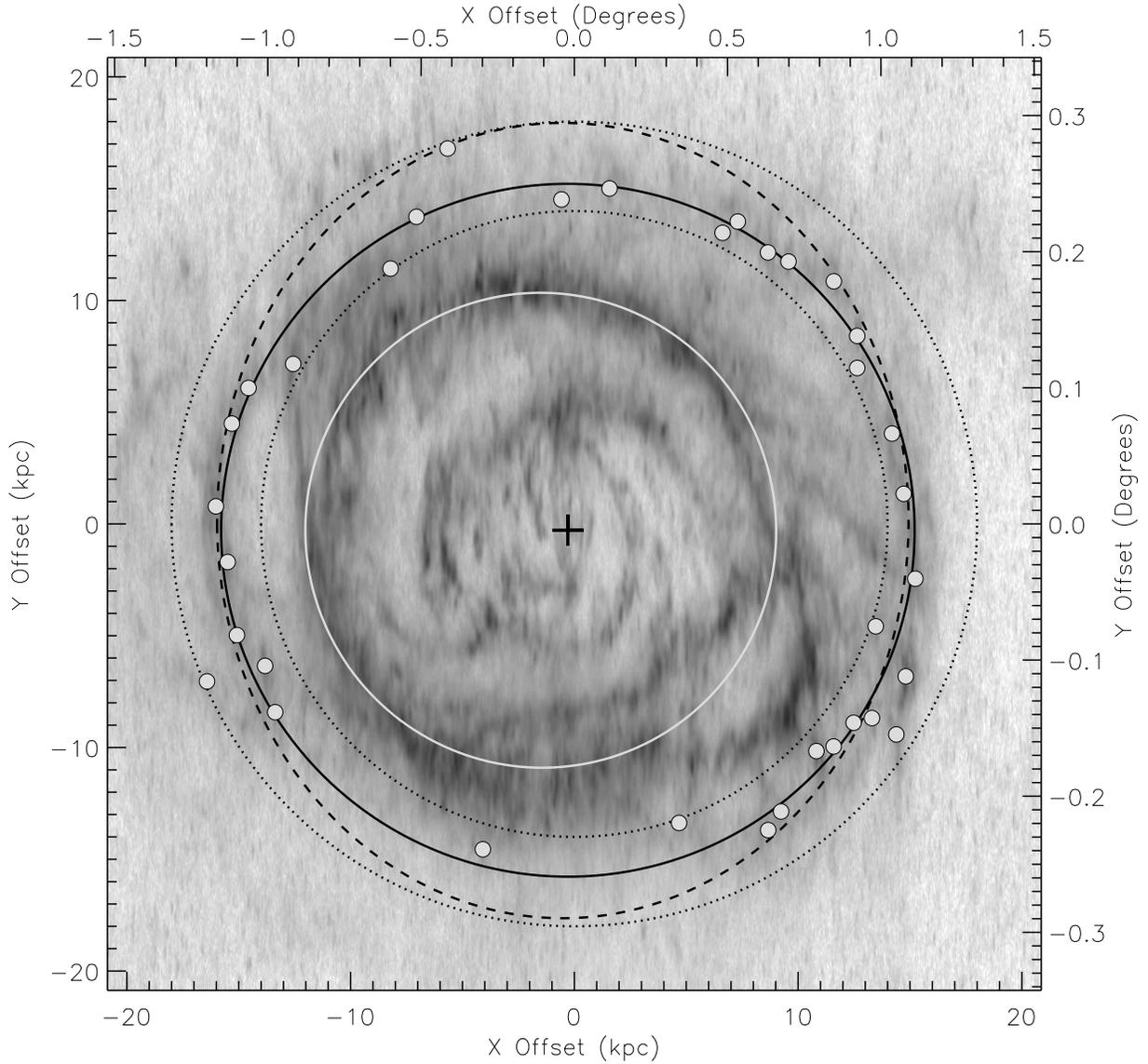}
	}
	\caption{\label{fig:15kpc} The 15\,kpc Ring (black). Greyscale shows deprojected SPIRE 350$\mu$m map (log scaling). The white Ring is the best fit to the 10\,kpc Ring. The white dots show clouds that were used in the 15\,kpc Ring fit and the dotted lines show the radii between which these were selected. The solid black line shows the best fit 15\,kpc Ring under the assumption of a $77^\circ$ inclination angle. The dashed line shows the equivalent under the assumption of a $75^\circ$ inclination angle. } 
\end{figure}

\begin{table}
	\caption{\label{tab:struct} Structural parameters for M31 by assumed angle of inclination.}
\centering{
	\begin{tabular}{ccccc}
	\hline\hline
	 Inclination & $R_{10\textrm{ kpc}}$ & $R_{15\textrm{ kpc}}$ & Arm Pitch Angle $\phi$ \\
	 Angle $i$ & & & \\
	($^\circ$) & (kpc) & (kpc) & ($^\circ$)  \\
	\hline
	77	& 10.5 & 15.5& 8.9  \\
	75	& 9.7 & 15.3 & 9.3  \\
	\hline
	\end{tabular}
}
\end{table}

Figure~\ref{fig:anterior} (left) shows a 350$\mu$m greyscale map of the centre of M31. The branch-network that comprises the 10\,kpc Ring (as identified by the box/contour in Figure \ref{fig:dendro}) is shown by the grey-contour. The position of these GMCs is shown by the circular markers. An offset circle is fit to them in two-stages: we exclude GMCs in the position angle (PA) range 120--240$^\circ$ (shown by the dotted wedge in Figure \ref{fig:anterior}). This gives an initial fit to the undistributed portion of the Ring. We then repeat the fit for all Ring GMCs over all PAs with a galactocentric radius within 1.5\,kpc of the first fit's result. These GMCs are shown by a cross inside their marker. It is the fit to these filtered GMCs that gives us the parameters for the Ring. 

The best-fit Ring is shown by the black circle on Figure \ref{fig:anterior}. It has a radius of $10.52\pm0.02$\,kpc and is offset from the assumed centre of M31 by 1.5\,kpc along the negative x-axis. The offset centre is shown by the cross. It has a Right Ascension of $0^{h}\,51^{m}\,18\fs 2$ and a Declination of $42\degr\,33\arcmin\,58\farcs0$. The exact extent of the deprojected ring is sensitive to M31's assumed angle of inclination. This angle can be estimated by fitting position-velocity tilted ring models to molecular line data \citep{2009ApJ...705.1395C,2010A&A...511A..89C} of M31. We have used the HELGA assumed  inclination angle of $77^\circ$ to fit our best fit radius of 10.5\,kpc. This angle is based on a mean value from the \citep{2009ApJ...705.1395C} model. However, in their original analysis \citet{2006ApJ...638L..87G} derived a radius of 9.8\,kpc using an inclination angle of $75^\circ$. This is closer to the mean value found in the \citet{2010A&A...511A..89C} model. We repeated our fitting using the \citet{2006ApJ...638L..87G} angle of inclination and found a radius of 9.7\,kpc in excellent agreement with their value. The 9.7\,kpc fit is shown by the dashed circle on Figure~\ref{fig:anterior} (it appears as an ellipse due to the differences in the assumed inclination angles). The differing results for the two inclination angles are listed in Table~\ref{tab:struct} and the differences between the two inclination models is discussed in Appendix \ref{deprojection}.

Figure~\ref{fig:anterior} (right) shows a radial profile of the 10\,kpc ring constructed in the coordinate frame of the 10.5\,kpc Ring fit. Normalised flux profiles for each of the 5 {\it Herschel} wavelengths are shown as solid lines, the colours are the same as for Figure~\ref{fig:slice}. The profiles were repeated with the exclusion of data in the PA range 120--240$^\circ$. The second set of profiles are shown by dashed lines and have been normalised to 0.8 as to offset them from the first set of profiles. All wavelengths longer than 350$\mu$m have been convolved to the 350$\mu$m resolution and pixel grid. 

There is a strong coincidence in the flux profiles on the interior side of the Ring, all reaching a minimum at 7.5\,kpc and a maximum at 10.5\,kpc. The correlation is particularly strong between 160--500$\mu$m suggesting that the cold dust component has a uniform temperature between these radii. The reverse is true on the outside of the ring where the long wavelength bands become increasingly strong. Comparison of this trend to the radial dust fits of \citetalias{2012arXiv1204.0785S} shows that there was only a slight temperature gradient in the outer galaxy. However, there was a stronger radial trend in the dust spectral index ($\beta$). A flattening of the Rayleigh-Jeans part of the dust SED, as shown by the changing $\beta$ profile \citetalias{2012arXiv1204.0785S}, could explain the divergence of the flux profiles seen in Figure~\ref{fig:anterior}.    

\subsection{The 15\,kpc Ring}

\begin{figure*}
	\centering{
		\includegraphics[width=\textwidth]{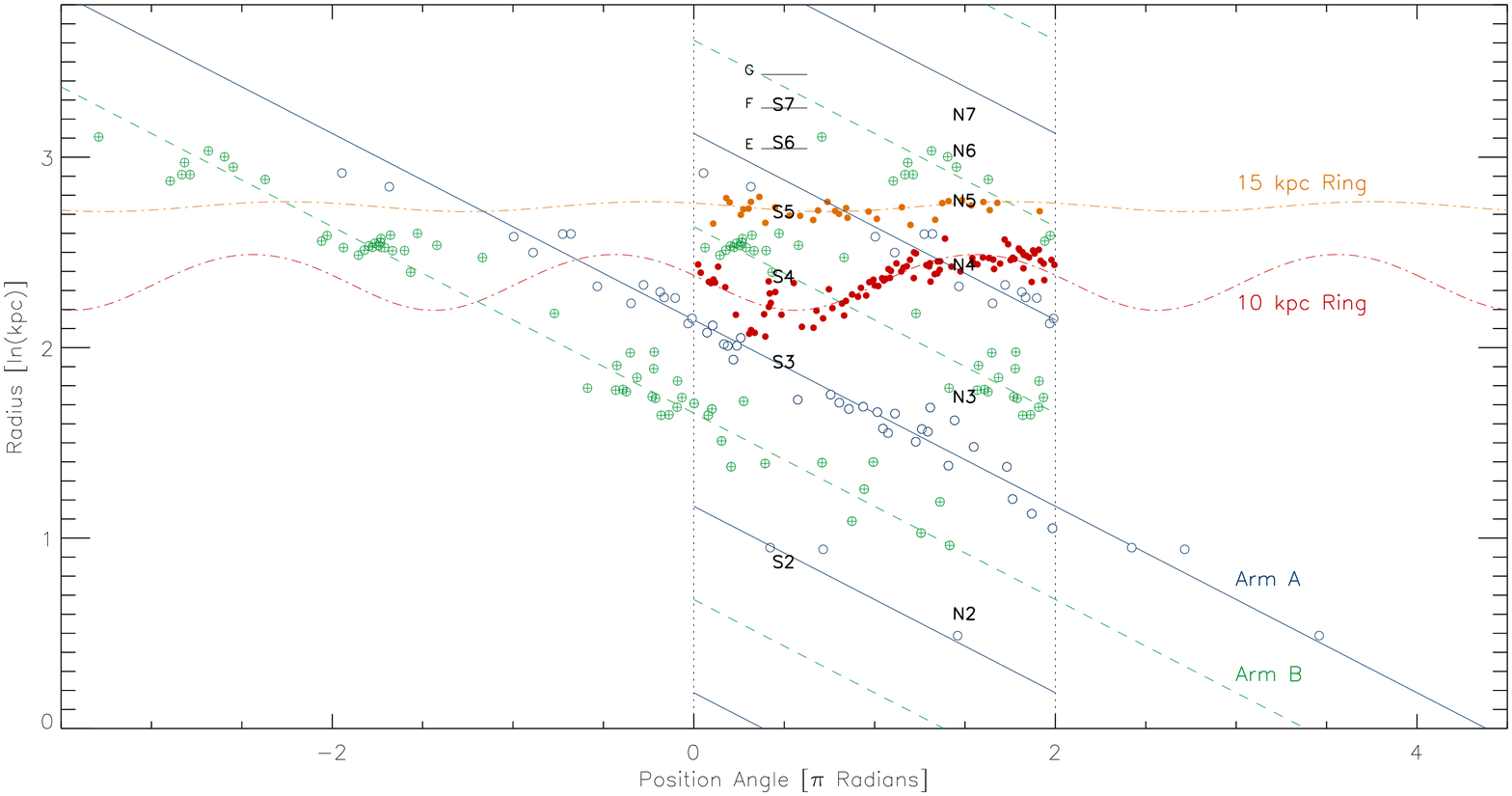}
	}
	\caption{\label{fig:polar} Polar coordinate plot of M31 GMCs. All GMCs are shown for a single 0-2$\pi$ radians range. GMCs used to fit the 10 or 15\,kpc Ring are shown by solid markers. The offset circle fits to the Rings are shown by the dot-dashed sinewaves. GMCs used to fit the spiral arms are shown by the open circles. The GMCs associated with Arm B are differentiated from those associated with Arm A by a cross inside their marker. The best-fit logarithmic spirals are shown by the solid line (Arm A) and the dashed line (Arm B). These Arm fits are replicated in the 0-2$\pi$ radians range to show how they wrap. Additionally, for each Arm we unwrap and replicate the GMCs associated with it to show the full fit. The arms have the same pitch angle and are a rotation of one another. The \citet{1963esag.book.....B} arm crossing regions are annotated. The E, F, and G features reported by  \citetalias{2011arXiv1112.3348F} are shown by the short horizontal lines. In the online version of this image the features associated with Arm A, Arm B, the 10\,kpc Ring, and the 15\,kpc Ring are respectively colour coded green, blue, red, and orange.  }
\end{figure*}

\begin{figure}
	\centering{
		\includegraphics[width=\columnwidth]{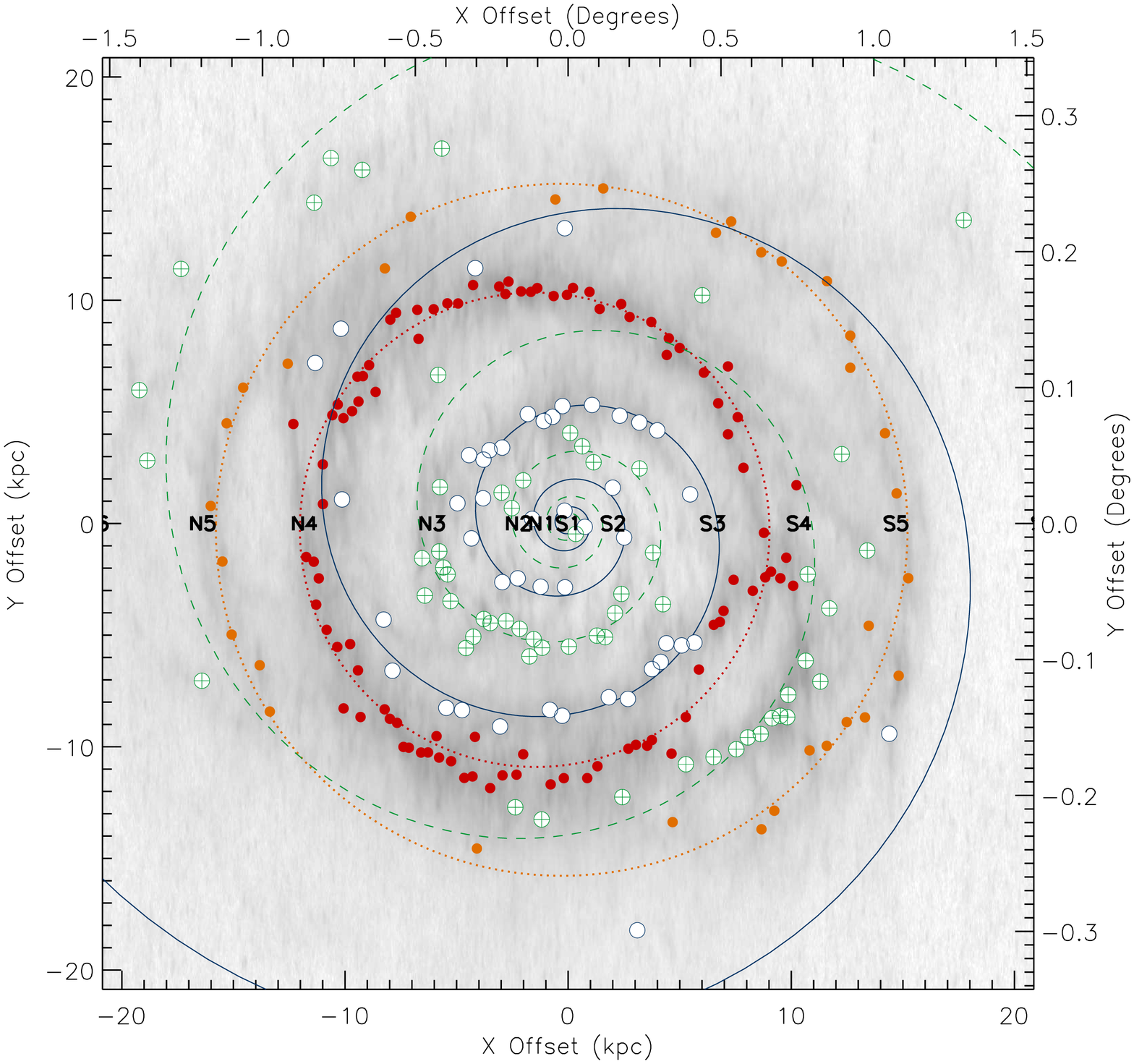}
	}
	\caption{\label{fig:arms} M31's spiral arms. The same region shown in Figure~\ref{fig:15kpc}. The 10 and 15\,kpc Rings are shown by dotted circles. GMCs associated with the arms are shown by open markers, GMCs associated with Arm B are differentiated by the cross over their marker. The best-fit logarithmic spirals are shown by the solid line (Arm A) and the dashed line (Arm B). The \citet{1963esag.book.....B} arm crossing regions are annotated. In the online version of this image the features associated with Arm A, Arm B, the 10\,kpc Ring, and the 15\,kpc Ring are respectively colour coded green, blue, red, and orange. }
\end{figure}

The existence of a 15\,kpc ring visible in infra-red maps of M31 has been noted before \citep{1998A&A...338L..33H,2006ApJ...638L..87G}. Features associated with this 15\,kpc ring are visible in the {\it Herschel} maps (e.g., Figure~\ref{fig:rgbproj}) and are seen as an enhancement in the number density of molecular clouds at that radii (see Figure ~\ref{fig:cloudprop2}). It is reasonable then to investigate whether this ring can be fit in the same manner as the 10\,kpc Ring. We do not have a unique tree branch for this structure as there is not a constant valley between it and the 10\,kpc Ring. To get around this we select all the GMCs with a galactocentric distance larger than 14\,kpc (to exclude the 10\,kpc ring and NGC 206) and less than 18\, kpc (to exclude the outer arcs) and fit them with an offset circle. 

Figure~\ref{fig:15kpc} shows a 350$\mu$m map of M31 scaled to show the structures at ${\sim}15$\,kpc. The two dashed lines show 14 and 18\,kpc bands used to filter the clouds. The best fit Ring radius of $15.50\pm0.02$\,kpc is shown by the solid line. The offset centre is shown by the cross. It has a Right Ascension of $0^{h}\,50^{m}\,58\fs 6$ and a Declination of $42\degr\,28\arcmin\,26\farcs0$. As with the 10\,kpc Ring, the fit is repeated for an inclination of $77^\circ$ and gives a best fit radii of 15.34\,kpc. The lower inclination fit is shown by the dash ellipse on Figure~\ref{fig:15kpc}. 

\citet{2006ApJ...638L..87G} investigated the offset of the 10\,kpc ring by modelling M31's interaction with the satellite galaxy M32. They showed that the passage of M32 could have triggered a wave of star formation which forms the ring itself. However, M32's proximity also causes the ring to be slightly pulled off-centre, thus creating the observed offset. The $xy$ offset we find for the 10\,kpc and 15,kpc rings are ( -1.5, -0.3 )\,kpc and ( -0.3, -0.3 ) kpc respectively. While different, both offsets do pull towards the negative-$x$, negative-$y$ direction suggesting that they are concentric with each other.

We stress that this exercise -- fitting a ring to set of circularly distributed points -- is only an aid to estimating the radius of feature seen in Figure~\ref{fig:rgbproj} and noted elsewhere \citep{1998A&A...338L..33H,2006ApJ...638L..87G}. Nevertheless, the 15.5\,kpc Ring does appear to closely match the distribution of GMCs. The reduced-$\chi^2$ value for this fit is ${\sim}0.8$ indicating that the fit is not unreasonable even if it is slightly over-constrained (as would be expected from the filtering). The $75^\circ$ fit gives a marginally higher reduced-$\chi^2$, but does not appear to fit the GMCs along the top of the ring (positive-$y$) so well.

\subsection{The Spiral Arms}

The ``spiral arms'' shown in Figure \ref{fig:rgbproj} are not contiguous and appear to be comprised of a series of disjointed arm segments. In order to better describe these arms we use the positions of the GMCs from our source extraction. These are filtered to remove GMCs that are within 1.5\,kpc of either of the fitted 10 or 15\,kpc Rings. We then use a preliminary by-eye fit to divide the remaining GMCs into those associated with one or other of two spiral arms. The two sets of GMCs appear to follow the same trend and suggest that they are merely a rotation of one another. The sources associated with each arm are shown plotted using polar coordinates in Figure~\ref{fig:polar} and using XY-offset coordinates in Figure~\ref{fig:arms}. 

For ease of reference we assign designations to each arm. We refer to the first arm as ``Arm A''. It is the arm that passes through the \citet{1963esag.book.....B} S3 and N4 arm crossing points. GMCs associated with Arm A are plotted as open circles. This arm appears to start around the S2 point and winds clockwise until it passes through the band of GMCs at $Y=+5$\,kpc, though the S3 point, and then through another bands of GMCs at $Y=-8$\,kpc close to the 10\,kpc Ring. We refer to the second arm as ``Arm B'' and plot the GMCs associated with it as circles with crosses. Arm B begins with a group of GMCs interior to Arm A at $Y=+3$\,kpc and then winds through a band of GMCs at $Y=-5$\,kpc before fading out as it passes through the N3 arm crossing region. GMCs associated with Arm B reappear over $180^\circ$ degrees later as it passes through the S4 point and picks up again as a strong spur of emission and clouds midway between the 10\,kpc and 15\,kpc Rings (NGC 206). Arm B finally passes through an elongated scatter of GMCs at $r>15$\,kpc in the top-left of Figure~\ref{fig:arms}. 

Figure~\ref{fig:polar} is split into two regions. The first region is wrapped over the range $0-2\pi$ radians (bracketed by the dotted lines) and is where all the GMCs are plotted with their position angles. Within this range the GMCs used to fit each of the 10 and 15\,kpc Rings are shown by the solid markers and the fits to the rings are shown by the dot-dashed lines. The equation for a classic logarithmic spiral can be written in the form $r = a \exp (b \theta)$ where $r$ and $\theta$ are the position of the spiral in polar coordinates, $a$ is a reference radius determining the relative rotation of the spiral, and $b$ is a constant related to the pitch angle $\phi$ by $b = 1 /\tan( 90 - \phi)$. Thus spiral arm segments will show up as linear features on a plot of $\ln(r)$ versus $\theta$. The errors on $a$ and $\phi$ are calculated from the fitted errors $\sigma_{\ln a}$ and $\sigma_b$ using the formulae $\sigma_{a} = \sigma_{\ln a} e^{a}$ and $\sigma_{\phi}^2 = \sigma_{b}^2 ( b^{2} + 1 )^{-2}$.

Figure~\ref{fig:polar} also shows the effect of linearising (removing the phase shift in the wrapped positions) of the Arm GMCs. Thus each arm GMC is plotted once in the $0-2\pi$ range and then once outside that range with the phase shift removed. A linear regression was preformed on the unwrapped Arm GMCs under the assumption that they had a common pitch angle and were offset from each other by $180^\circ$. The fit gave linear constants of $\ln(a)=1.90\pm0.01$\,$\ln(\textrm{\,kpc})$ and $b=-0.157\pm0.002$\,rad$^{-1}$\,$\ln(\textrm{\,kpc})$ which thus gives $a=6.7\pm0.1$\,kpc and $\phi=8.9\pm0.1^\circ$. Single fits to just Arm A and Arm B gives $\phi_{A}=9.0\pm0.2^\circ$ and $\phi_{B}=8.7\pm0.2^\circ$. 

As before, the fit was repeated using an assumed inclination angle of 75$^\circ$ and found a pitch angle of 9.3$^\circ$ in agreement with the pitch angles of 9.0$^\circ$ and 9.5$^\circ$ found by \citet{2006ApJ...638L..87G}. We also repeated the fit (for the original 77$^\circ$ inclination) without excluding the 15\,kpc Ring GMCs and found that the best-fit pitch angle was $8.8\pm0.1^\circ$, within the error bar of the original fit. The pitch angle for the spiral arms appears to be independent, or at least weakly influenced, by the sources associated with the 15\,kpc Ring. An estimate of ${\sim}9^\circ$ for the pitch angle of M31's spiral arms therefore appears reasonable given common assumptions for the inclination angle and the possible inclusion of features associated with the 15\,kpc Ring.  

Also plotted on Figure~\ref{fig:polar} are the positions of the \citet{1963esag.book.....B} arm crossing regions and the features E, F, and G from \citetalias{2011arXiv1112.3348F}. From our fit we see that regions S2 and S3 are associated with Arm A. However, neither N2 or N3 appear to be closely associated with either arm. The S4 region is associated with Arm B  while region N4 appears to be associated with the place when Arm A crosses the 10\,kpc Ring. Regions S5 and N5 are not near the fitted Arms and appear to be associated with the 15\,kpc Ring. The correspondence between the fit and features becomes more uncertain outside of the 15\,kpc Ring. Arm B does pass through a loose scatter of GMCs and through feature G. Features E and F are associated with regions S6 and S7, but neither N6, N7, S6 or S7 appear to be closely associated with any particular arm. It is possible that evolution in the Arm's pitch angle or changes in inclination angle are influencing features this far out. We explore the effects of a non-uniform inclination angle on M31's spiral structure in Appendix~\ref{deprojection}     

There is a disrupted portion of the Ring at $X=8, Y=-8$\,kpc that appears to be a inter-arm hole cleared out between the two spiral arms. \citet{2006ApJ...638L..87G} simulated the interaction of M32 with M31's disc and showed that such a hole could be created by M32's passage through the disc. The survival of this hole against differential rotation implies a relatively short timescale, on the order of 20 Myr \citep{2006ApJ...638L..87G}. A general effect of M32's passage could have been a wave of star formation within M31's disc \citep{2006ApJ...638L..87G}. M32's H{\sc ii} luminosity function is double peaked with the fainter peak being consistent with emission from a population of the B stars with a lifetime of 15 Myr \citep{2011AJ....142..139A}, comparable to the timescale for the passage of M32 through M31 disc.


\section{Summary}

We have used HELGA \citepalias{2011arXiv1112.3348F} data taken with the {\it Herschel Space Observatory} (wavelengths 100-500$\mu$m) to create a catalogue of molecular clouds in the nearby galaxy of Andromeda.

\begin{itemize}
	
	\item Monochromatic source extraction was performed on the M31 field using the hierarchical source extraction algorithm \textsc{csar} \citep{2013MNRAS.tmp.1218K}. The tree of sources was pruned back to that containing the contiguous emission from M31 alone. A total of 471 nodes were found in the structure tree. Of these, 236 were left-nodes, i.e., sources without resolved substructure. These are the sources that form the catalogue presented in this paper.
	
	\item The surface number density of clouds peaks towards the centre of M31 and falls off at a rate similar to that of the optical surface brightness out to 15\,kpc. On top of this distribution are a series of peaks at ${\sim}5$, 10, and 15\,kpc coincident with the reported rings of emission at several of those wavelengths. In addition, \citetalias{2011arXiv1112.3348F} found a series of arc-like features at ${\sim}20$, 25, and 30\,kpc suggesting that M31 contains a set of nested weak resonant rings whose radii are multiples of 5\,kpc.
	
	\item {\it Herschel} photometry was performed for each of the clouds. The temperature and mass of each cloud was found by fitting a greybody to its SED. The dust parameters were described by the radial dust relationships (dust-to-mass ratio, dust emissivity) from \citetalias{2012arXiv1204.0785S}. The median dust temperature was 18\,K. 
	
	\item Clouds with masses in the range $10^4-10^7$\,M$_\odot$ and with sizes of ~100-1000\,pc were found. This and a comparison with interferometry maps showed that we are resolving structures that are comparable to large GMCs and complexes of multiple GMCs within the Milky Way. The powerlaw slope of the cloud’s cumulative mass function agreed with that found in interferometric studies of M31.
	
	\item The clouds' properties appear to be consistent with those of clouds found in the Milky Way. Specifically, the far-infrared luminosity function, the relationship of far to mid-infrared luminosity, and the relationship of cloud mass to $^{12}$CO luminosity are all consistent with that found for clouds in the Milky Way. The last relationship was found to virtually identical to that found by \citet{1987ApJ...319..730S} for clouds in the Milky Way.
	
	\item Following, \citet{2006ApJ...638L..87G} we fit an offset circle to the dominant ring feature and calculate a radius of 10.5\,kpc. Our results were consistent with \citet{2006ApJ...638L..87G}, allowing for differences in assumed inclination angle.  We also fit an offset circle to the clouds at ~15\,kpc and derive radius of 15.5\,kpc. The centres of both Rings are offset in the same approximate direction from the assumed centre of M31.  
	
	\item Clouds associated with the 10 and 15\,kpc Rings were excluded and a logarithmic spiral was fit to the remaining sources. A common pitch angle of $8.9^\circ$ was found for two spiral arms that trailed one another by $180^\circ$. The fitted arms and rings are consistent with the arm crossing features described by \citet{1963esag.book.....B}.

\end{itemize}

\acknowledgments

SPIRE has been developed by a consortium of institutes led by Cardiff University (UK) and including Univ. Lethbridge (Canada); NAOC (China); CEA, LAM (France); IFSI, Univ. Padua (Italy); IAC (Spain); Stockholm Observatory (Sweden); Imperial College London, RAL, UCL-MSSL, UKATC, Univ. Sussex (UK); and Caltech, JPL, NHSC, Univ. Colorado (USA). This development has been supported by national funding agencies: CSA (Canada); NAOC (China); CEA, CNES, CNRS (France); ASI (Italy); MCINN (Spain); Stockholm Observatory (Sweden); STFC (UK); and NASA (USA).

PACS has been developed by a consortium of institutes led by MPE (Germany) and including UVIE (Austria); KU Leuven, CSL, IMEC (Belgium); CEA, LAM (France); MPIA (Germany); INAFIFSI/OAA/OAP/OAT, LENS, SISSA (Italy); IAC (Spain). This development has been supported by the funding agencies BMVIT (Austria), ESA-PRODEX (Belgium), CEA/CNES 
(France), DLR (Germany), ASI/INAF (Italy), and CICYT/MCYT (Spain).

\appendix

\section{Deprojection}
\label{deprojection}

\begin{figure*}
\centering{
	\includegraphics[width=0.5\textwidth]{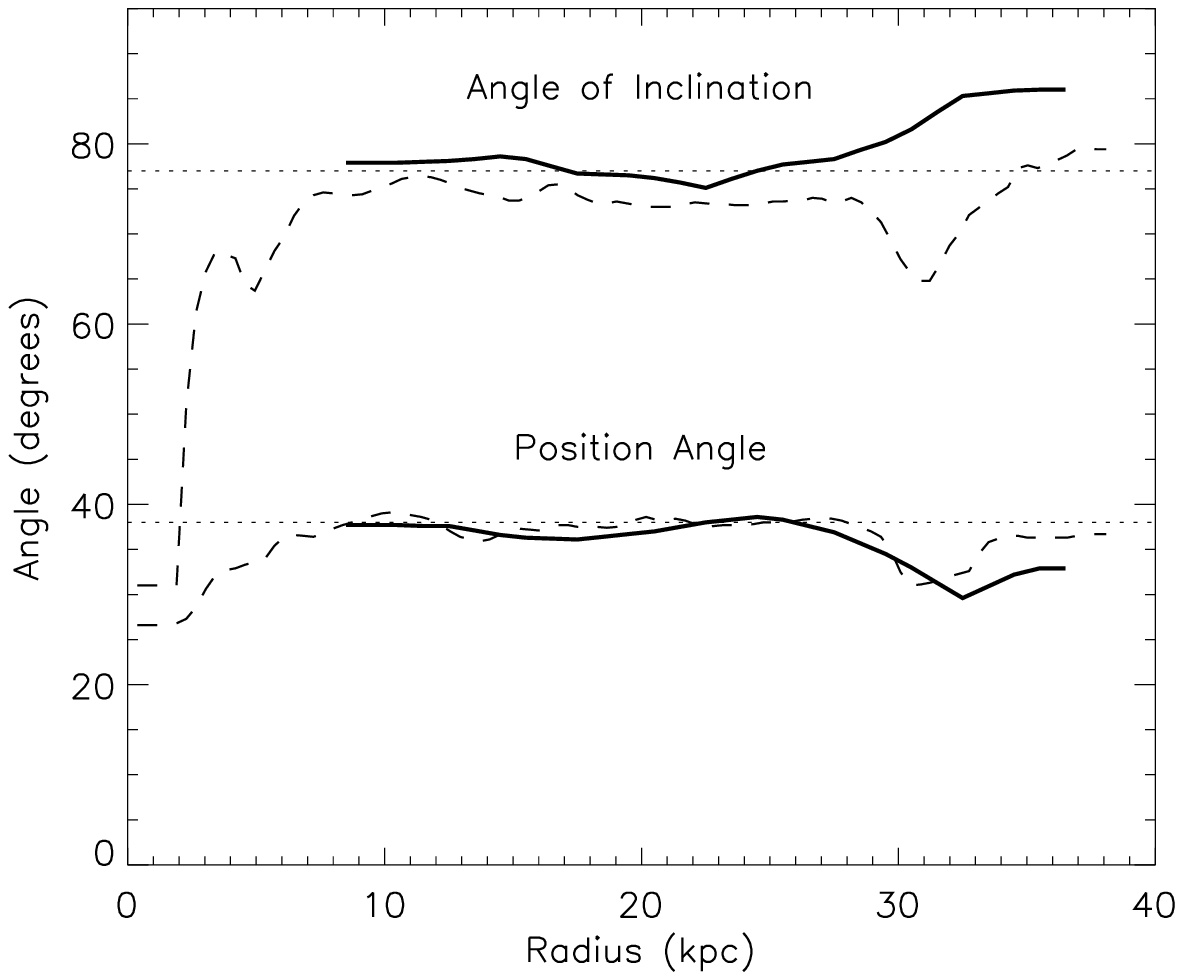}
	}
\caption{\label{fig:radmodels}Inclination angles (top) and position angles (lower) from the Chemin (dashed-lines) and Corbelli (solid-lines) tilted-ring models. The dotted-lines show the values adopted by HELGA.  }
\end{figure*}

\begin{figure*}
\centering{
	\includegraphics[width=\textwidth]{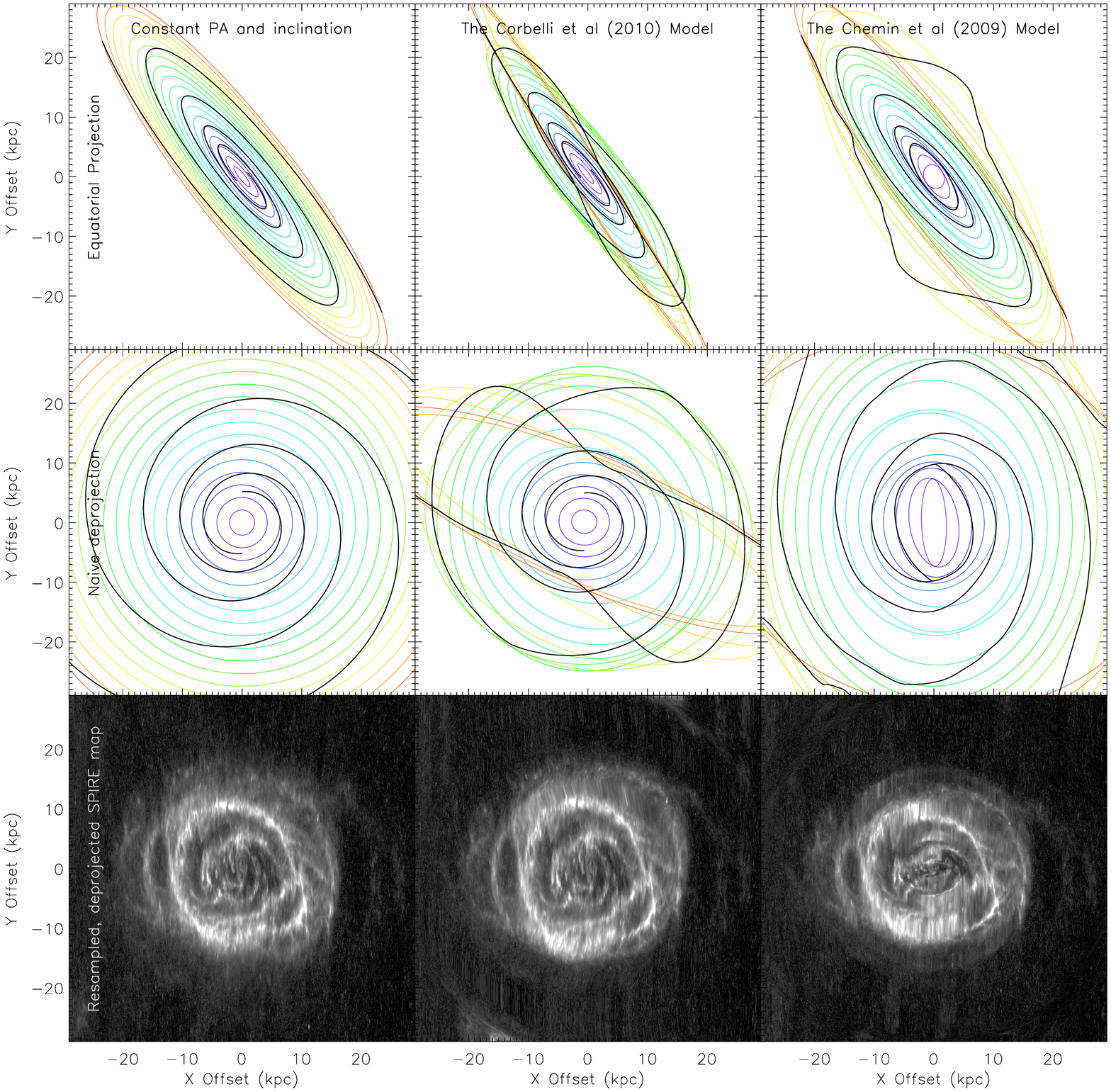}
	}

\caption{\label{fig:dpmod} The effects of deprojection scheme on structure in M31. A toy model of concentric rings (coloured circles) overlain by twin-spiral arms (black curves) is taken as the initial conditions. The top row shows the toy model projected into equatorial equivalent offsets, the middle row shows the equatorial equivalent projection deprojected under the assumption of constant $i$ and $\theta$, and the bottom row shows the deprojected SPIRE 350$\mu$m maps. Three projection models are shown: the left-hand column shows the results of using constant $i$ and $\theta$, the middle column shows the Corbelli Model, and the right-hand column shows the Chemin Model. }
\end{figure*}

It has been known since the earliest studies of Andromeda's structure \citep{1963esag.book.....B,1964ApJ...139.1045A} that its stellar disc exhibited a pronounced warp at large radii. This warp means that studies which use a constant position and inclination angle to deproject M31 are liable to introduce artefacts resulting from differences between the assumed flat geometry and the actual warped geometry. In order to quantify possible problems of this sort we study the effects of projecting M31 using two recent models published by \citet[][hereafter the Chemin Model]{2009ApJ...705.1395C} and \citet[][hereafter the Corbelli Model]{2010A&A...511A..89C} using independent H{\sc i} surveys. Direct comparison of the models is made easier as both use the same distance to Andromeda as adopted by the HELGA consortium \citep{2005mcconnachie}.

The two literature models analysed M31 as a series of nested, tilted rings. Each ring represents the projection of a particular circular orbit with its own angle of inclination $i(R)$ and position angle $\theta(R)$. Figure \ref{fig:radmodels} plots the inclination and position angles from table 4 of \citet{2009ApJ...705.1395C} and table 1 of \citet{2010A&A...511A..89C}. The Chemin Model tabulated parameters for the entire disc from the centre to 38\,kpc, but the Corbelli Model only tabulated values over the range 8.5--36.5\,kpc as they did not model the inner part of the H{\sc i} disc. The Corbelli Model includes small offsets ($x_0, y_0$) of order 1-2 arcmin (less than 0.5\,kpc) to the central position of each ring, but the Chemin Model does not.     

Figure \ref{fig:radmodels} shows that the position angles adopted by the two models broadly agree. However, the angle of inclination adopted by the Corbelli Model is systematically higher than the value adopted by the Chemin Model. The mean inclination angle over the range 10-20\,kpc is $75\pm1^\circ$ for the Chemin Model and $77\pm1.0^\circ$ for the Corbelli Model (this is value assumed by the HELGA survey) . Likewise, the mean position angle over the same range is $37.5\pm0.9^\circ$ for the Chemin Model and $37.3\pm0.8^\circ$ for the Corbelli Model. These inclination angles represent a deprojection factor of $\times3.86$ for the Chemin Model and $\times4.44$ for the Corbelli Model along the projected minor axis, a difference of approximately 15\%.  The models also differ in their trends in the outer disc, where the warp should be most noticeable. Both Models include a position angle minimum at around $\sim32$\,kpc. The Chemin model includes a similar drop in the inclination angle, but the Corbelli Model does not and has the inclination angle increasing monotonically in the outer disc.  

To examine the effects of these two models we construct a simple toy model of M31 consisting of a series of concentric circular rings and twin logarithmic spiral arms with a pitch angle of $8.5^\circ$. The top row of Figure \ref{fig:dpmod} shows the effects of projecting the toy model on to the plane of the sky using the constant angles assumed by HELGA ($i=77^\circ$, $\theta=38^\circ$), the Corbelli Model, and the Chemin Model. For the flat (constant angles) geometry the inclination of the disc means that the rings along the projected minor-axis become very close, but never over lap. By comparison, after $\sim27$\,kpc, the Corbelli and Chemin Models deviate from the approximately linear trend of position angle with radius. This causes the rings to precess against one another and to overlap. 

The position angle trend is amplified by the aforementioned divergence of the adopted inclination angle. The effect of this is to send the outer parts of the spiral arms in opposite directions. The Corbelli Model causes the outer rings, and the outer spiral arm segments, to be projected inwards over/below the central part of the galaxy. By contrast the Chemin Model causes those same spiral arms to flare outwards along the minor axis. There are also differences in the centre of the galaxy as the Chemin Model causes the inner part of the spiral arms to merge into a ring like structure. No projection data was give for the Corbelli Model within 8.5\,kpc so we used the data from that radius for the interior portion. 

The effects of naively deprojecting warped structures whilst using a non-warped assumption are explored in the middle row of Figure \ref{fig:dpmod}. Here the projected models from the top row are deprojected using the constant position and inclination angles used for the first column. As expected, the first column deprojects perfectly, but there are significant artefacts introduced into the other two panels. The assumed $i$ and $\theta$ most closely match the Corbelli Model so it is unsurprising that it appears the most circular. This exercise reinforces how structure assumed to be at one radius, particularly faint structure as would be found on the tail end of a spiral arm, may actually be at a completely different radius. 

The bottom line of panels shows the 350$\mu$m map of M31 (the one used for source identification) resampled into a rectilinear face-on-grid under the assumptions of the flat geometry, the Corbelli, and the Chemin Models. These maps were created by calculating the Right Ascension and Declination for every pixel under the assumptions of each model. Each pixel was then assigned the brightness of the original map at that R.A. and Dec. For the constant angle example each $x,y$ pixel mapped uniquely onto a single R.A., Dec position. However, there was a degeneracy in the tilted ring models   where multiple $x,y$ pixels mapped onto the same R.A., Dec position, as would be expected from the overlapping rings in the preceding panels. This effect created regions on the deprojected maps where features were stretched out and blurred. 

Comparison of the deprojected 350$\mu$m map in Figure \ref{fig:dpmod} with the other panels shows that the majority of emission is within a radius of $\sim22$\,kpc and is this not directly effected by the strongest parts of the outer warp. Indeed, the warp only becomes important when the Corbelli Model projects spiral arms over the centre of the galaxy. Of the two variable $i$, $\theta$ models it is the Chemin Model which gives a version of M31 that appears the most circular at large radii. However, it also displays significant degeneracy along the projected short axis. It does, unlike the Corbelli Model, deproject the centre of the galaxy. The lozenge/bar shaped inner structure shown in the middle-right panel of Figure \ref{fig:dpmod} is similar to that seen in the RGB image shown in Figure \ref{fig:rgb}.


\section{Satellites}

\begin{figure}
	\centering{
		\includegraphics[width=0.8\textwidth]{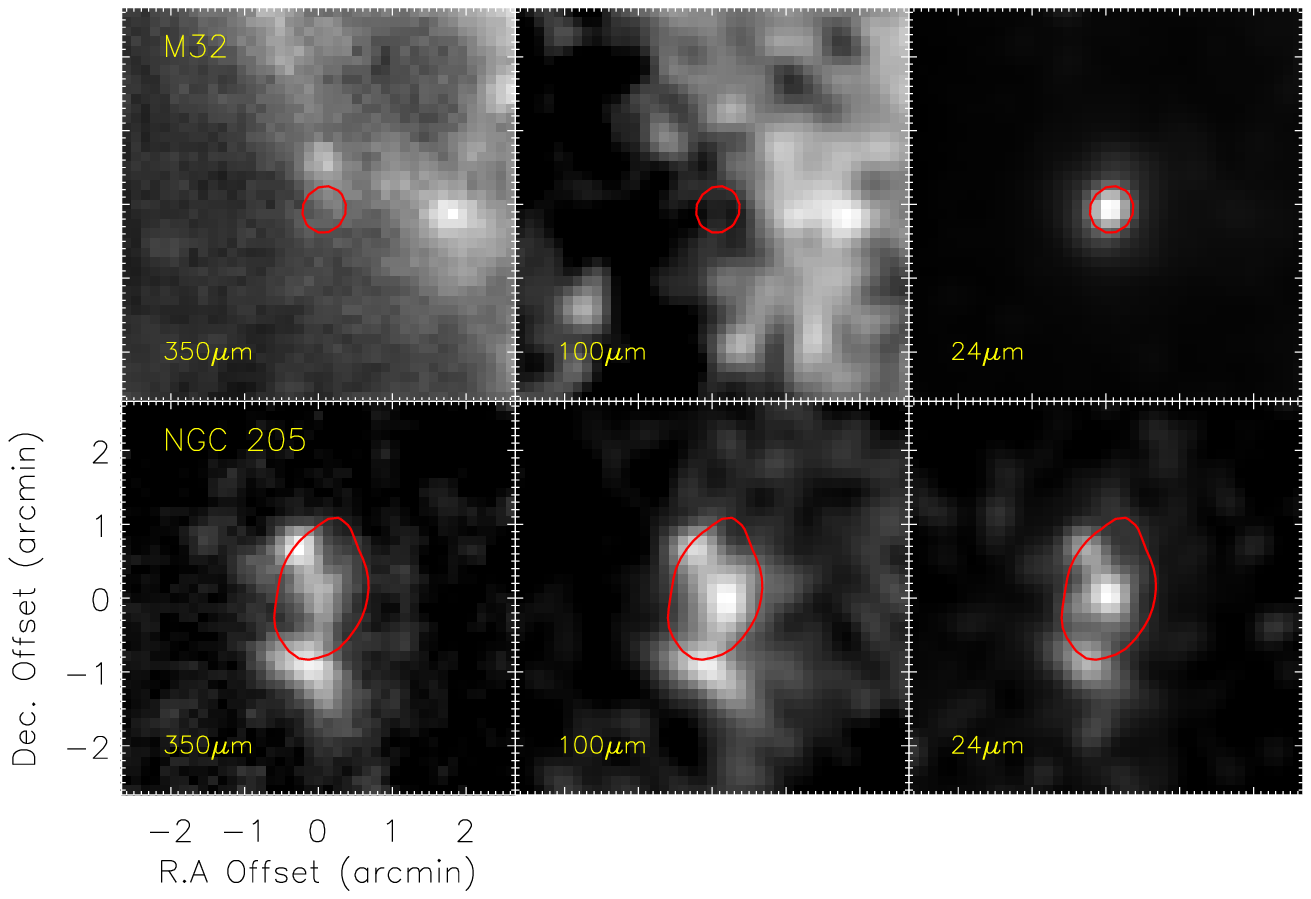}
	}
	\caption{\label{fig:sat} Two of the satellites of Andromeda, M32 and NGC 205, showing markedly different emission properties with wavelength. The individual wavelengths are labelled. The data has been convolved to the 350$\mu$m resolution. The IRAC 3.6$\mu$m 50\% flux contour is shown in red. The position centre for M32 is $0^{h}\,42^{m}\,41\fs 87$  $40\degr\,51\arcmin\,57.2\farcs50$ and for NGC 205 is $0^{h}\,40^{m}\,00\fs 08$  $41\degr\,41\arcmin\,07.1\farcs50$.} 
\end{figure}

There are two dwarf galaxies within the M31 field, M32 and NGC 205, which are of note for the dramatically varying dust emission. Figure \ref{fig:sat} shows both galaxies at SPIRE 350$\mu$m, PACS 100$\mu$m, and {\it Spitzer} MIPS 24$\mu$m, the contour shows the {\it Spitzer} IRAC 3.6$\mu$m 50\% peak intensity contour. All the data has been convolved to the 350$\mu$m resolution. M32 appears strongly at 24$\mu$m, but is completely devoid of emission in the {\it Herschel} images. By contrast, NGC 205 shows revolved emission at three positions - a central peak coincident with short wavelength centre and long wavelengths peaks to the north and south of it. See \citet{2012MNRAS.423.2359D} for a study of the NGC 205 data.

\end{document}

%% file: Andromeda_350res_table1_narrow_short.tex
\begin{deluxetable}{ c l l cccc cccc}
\tablecaption{\label{tab:leafs}Table of measured leaf-node parameters. }
\tablecolumns{11}
\tabletypesize{\scriptsize}
\tablewidth{0pt}
\rotate
\tablehead{ 
\colhead{Name} & \colhead{R.A. (2000)} & \colhead{Dec. (2000)} & \colhead{$X$\tablenotemark{a}} & \colhead{$Y$\tablenotemark{a}} & \colhead{$R$\tablenotemark{a}} & \colhead{FWHM\tablenotemark{b}} & \colhead{$S_{350\mu m}$} & 	\colhead{$S_{250\mu m}$}  & \colhead{$S_{160\mu m}$}  & \colhead{$S_{100\mu m}$}  \\
 (HELGA) &  &  &  [kpc] & [kpc] & [kpc] & [pc] & [Jy] & [Jy] & [Jy] & [Jy]  \\
}
\startdata
       1 & $00^{\mathrm{h}}$ $46^{\mathrm{m}}$ $59\fs9$ & $42\degr$ $25\arcmin$ $44\arcsec$ & -19.2 &   6.0 &  20.1 &      1400. & $    5.0\pm    0.1$ & $    5.7\pm    0.1$ & $    4.1\pm    0.4$ & $    1.8\pm    0.3$ \\
       2 & $00^{\mathrm{h}}$ $47^{\mathrm{m}}$ $07\fs7$ & $42\degr$ $22\arcmin$ $34\arcsec$ & -18.9 &   2.8 &  19.1 &      78. & $  0.082\pm  0.004$ & $  0.087\pm  0.008$ & $<  0.095$ & $<  0.093$ \\
       3 & $00^{\mathrm{h}}$ $39^{\mathrm{m}}$ $18\fs7$ & $40\degr$ $21\arcmin$ $51\arcsec$ &  15.2 &  -2.4 &  15.4 &      92. & $    2.1\pm    0.1$ & $    3.7\pm    0.1$ & $    4.7\pm    0.2$ & $    3.2\pm    0.2$ \\
       4 & $00^{\mathrm{h}}$ $39^{\mathrm{m}}$ $10\fs3$ & $40\degr$ $25\arcmin$ $56\arcsec$ &  14.7 &   1.3 &  14.8 &      290. & $  0.35\pm  0.01$ & $  0.54\pm  0.02$ & $  0.47\pm  0.05$ & $  0.15\pm  0.03$ \\
       5 & $00^{\mathrm{h}}$ $39^{\mathrm{m}}$ $06\fs5$ & $40\degr$ $29\arcmin$ $21\arcsec$ &  14.2 &   4.0 &  14.8 &      290. & $  0.48\pm  0.01$ & $  0.79\pm  0.03$ & $  0.64\pm  0.09$ & $  0.23\pm  0.06$ \\
       6 & $00^{\mathrm{h}}$ $39^{\mathrm{m}}$ $42\fs6$ & $40\degr$ $20\arcmin$ $43\arcsec$ &  14.8 &  -6.8 &  16.3 &      300. & $    6.1\pm    0.1$ & $     11.\pm      1.$ & $     16.\pm      1.$ & $    8.8\pm    0.3$ \\
       7 & $00^{\mathrm{h}}$ $39^{\mathrm{m}}$ $59\fs1$ & $40\degr$ $20\arcmin$ $38\arcsec$ &  14.4 &  -9.4 &  17.2 &      220. & $  0.58\pm  0.02$ & $  0.92\pm  0.04$ & $  0.68\pm  0.09$ & $  0.26\pm  0.07$ \\
       8 & $00^{\mathrm{h}}$ $39^{\mathrm{m}}$ $39\fs3$ & $40\degr$ $28\arcmin$ $59\arcsec$ &  13.4 &  -1.2 &  13.5 &      190. & $    3.9\pm    0.1$ & $    6.9\pm    0.1$ & $    7.7\pm    0.3$ & $    3.2\pm    0.2$ \\
       9 & $00^{\mathrm{h}}$ $45^{\mathrm{m}}$ $36\fs6$ & $41\degr$ $54\arcmin$ $16\arcsec$ & -11.4 &  -1.7 &  11.5 &      110. & $  0.20\pm  0.01$ & $  0.41\pm  0.03$ & $  0.36\pm  0.05$ & $  0.26\pm  0.05$ \\
      10 & $00^{\mathrm{h}}$ $45^{\mathrm{m}}$ $36\fs4$ & $41\degr$ $53\arcmin$ $03\arcsec$ & -11.2 &  -2.5 &  11.4 &      78. & $  0.054\pm  0.007$ & $  0.11\pm  0.01$ & $<  0.11$ & $  0.082\pm  0.021$ \\
      11 & $00^{\mathrm{h}}$ $45^{\mathrm{m}}$ $40\fs7$ & $41\degr$ $55\arcmin$ $34\arcsec$ & -11.7 &  -1.5 &  11.8 &      150. & $    1.0\pm    0.1$ & $    1.9\pm    0.1$ & $    2.4\pm    0.2$ & $    1.1\pm    0.1$ \\
      12 & $00^{\mathrm{h}}$ $45^{\mathrm{m}}$ $43\fs0$ & $41\degr$ $52\arcmin$ $43\arcsec$ & -11.3 &  -3.6 &  11.9 &      140. & $  0.37\pm  0.02$ & $  0.75\pm  0.05$ & $  0.94\pm  0.14$ & $  0.87\pm  0.11$ \\
      13 & $00^{\mathrm{h}}$ $45^{\mathrm{m}}$ $37\fs0$ & $41\degr$ $48\arcmin$ $19\arcsec$ & -10.3 &  -5.5 &  11.7 &      160. & $  0.75\pm  0.03$ & $    1.5\pm    0.1$ & $    1.6\pm    0.3$ & $  0.85\pm  0.21$ \\
      14 & $00^{\mathrm{h}}$ $45^{\mathrm{m}}$ $40\fs8$ & $41\degr$ $50\arcmin$ $26\arcsec$ & -10.8 &  -4.8 &  11.8 &      130. & $  0.34\pm  0.02$ & $  0.71\pm  0.04$ & $  0.85\pm  0.12$ & $  0.40\pm  0.06$ \\
      15 & $00^{\mathrm{h}}$ $45^{\mathrm{m}}$ $27\fs8$ & $41\degr$ $44\arcmin$ $30\arcsec$ &  -9.4 &  -6.6 &  11.5 &      310. & $    2.3\pm    0.1$ & $    4.6\pm    0.1$ & $    6.0\pm    0.2$ & $    2.8\pm    0.2$ \\
      16 & $00^{\mathrm{h}}$ $45^{\mathrm{m}}$ $28\fs1$ & $41\degr$ $46\arcmin$ $26\arcsec$ &  -9.8 &  -5.4 &  11.2 &      78. & $  0.093\pm  0.012$ & $  0.18\pm  0.03$ & $  0.20\pm  0.05$ & $  0.17\pm  0.03$ \\
      17 & $00^{\mathrm{h}}$ $44^{\mathrm{m}}$ $55\fs8$ & $41\degr$ $29\arcmin$ $22\arcsec$ &  -5.9 & -11.4 &  12.8 &      78. & $  0.23\pm  0.01$ & $  0.53\pm  0.03$ & $  0.80\pm  0.09$ & $  0.49\pm  0.07$ \\
      18 & $00^{\mathrm{h}}$ $44^{\mathrm{m}}$ $51\fs1$ & $41\degr$ $29\arcmin$ $38\arcsec$ &  -5.8 & -10.5 &  12.0 &      78. & $  0.10\pm  0.01$ & $  0.24\pm  0.02$ & $  0.34\pm  0.04$ & $<  0.14$ \\
      19 & $00^{\mathrm{h}}$ $44^{\mathrm{m}}$ $43\fs9$ & $41\degr$ $27\arcmin$ $41\arcsec$ &  -5.2 & -10.6 &  11.9 &      210. & $    4.1\pm    0.1$ & $    8.7\pm    0.1$ & $     13.\pm      1.$ & $    6.8\pm    0.3$ \\
      20 & $00^{\mathrm{h}}$ $44^{\mathrm{m}}$ $32\fs8$ & $41\degr$ $23\arcmin$ $59\arcsec$ &  -4.3 & -11.3 &  12.1 &      140. & $  0.73\pm  0.02$ & $    1.8\pm    0.1$ & $    2.6\pm    0.2$ & $    1.2\pm    0.2$ \\
\enddata
\tablenotetext{a}{The source positions are calculated in the rotated deprojected frame ($\theta=38^\circ$, $i=77^\circ$) and are given relative to the assumed position centre.}
\tablenotetext{b}{The geometric mean of the deconvolved major and minor FWHM in the original non-rotated, projected frame.} 
\tablecomments{This table is published in its entirety in the electronic edition of this journal. A portion is shown here for guidance regarding its form and content.}
\end{deluxetable}

%% file: Andromeda_350res_table2_narrow_short.tex
\begin{deluxetable}{ c r@{.}l c r@{.}l r@{.}l }
\tablecaption{\label{tab:leafprops}Table of SED parameters. }
\tablecolumns{8}
\tablewidth{0pt}
\tablehead{ 
\colhead{Name} & \multicolumn{2}{c}{$M_\mathrm{cloud}$} & \colhead{$T$} & \multicolumn{2}{c}{$L_{FIR}$} & \multicolumn{2}{c}{$L_{\mathrm{CO}}$} \\
 (HELGA) &  \multicolumn{2}{c}{ [$10^5$\,M$_\odot$] } & [K] & \multicolumn{2}{c}{ [$10^5$\,L$_\odot$] } & \multicolumn{2}{c}{ [$10^4$\,K\,km\,s$^{-1}$ pc$^2$ ] } \\
}
\startdata
       1 &      160 &  &      17. &      22 &   & \multicolumn{2}{c}{ \nodata}   \\
       3 &      26 &  &      20. &      24 &   & \multicolumn{2}{c}{ \nodata}   \\
       4 &     8 & 0 &      16. &     2 & 0 & \multicolumn{2}{c}{ \nodata}   \\
       5 &      11 &  &      16. &     2 & 9 & \multicolumn{2}{c}{ \nodata}   \\
       6 &      88 &  &      21. &      71 &   & \multicolumn{2}{c}{ \nodata}   \\
       7 &      16 &  &      16. &     3 & 3 & \multicolumn{2}{c}{ \nodata}   \\
       8 &      60 &  &      17. &      32 &   & \multicolumn{2}{c}{ \nodata}   \\
       9 &     2 & 3 &      18. &     2 & 0 &     2 & 4 \\
      10 &   0 & 55 &      19. &   0 & 67 & $<$  0 & 52 \\
      11 &      12 &  &      18. &     9 & 8 &      14 &   \\
      12 &     2 & 9 &      21. &     5 & 7 &     3 & 6 \\
      13 &     9 & 3 &      18. &     7 & 3 &     7 & 2 \\
      14 &     4 & 1 &      18. &     3 & 6 & $<$    3 & 6 \\
      15 &      27 &  &      18. &      25 &   &      43 &   \\
      16 &   0 & 80 &      19. &     1 & 2 &   0 & 92 \\
      17 &     2 & 3 &      21. &     3 & 6 &     3 & 7 \\
      18 &   0 & 94 &      20. &     1 & 7 &     1 & 8 \\
      19 &      43 &  &      19. &      54 &   &      42 &   \\
      20 &     8 & 2 &      19. &      11 &   &      13 &   \\
      21 &     1 & 7 &      23. &     6 & 2 & $<$    1 & 3 \\
\enddata
\tablecomments{This table is published in its entirety in the electronic edition of this journal. A portion is shown here for guidance regarding its form and content.}
\end{deluxetable}